

\magnification\magstephalf
\baselineskip14pt
\vsize23.5truecm 
\hsize14.0truecm
\hoffset1.25truecm 





\def\hatt{\widehat}
\def\dell{\partial}
\def\tilda{\widetilde}
\def\eps{\varepsilon}

\def\half{\hbox{$1\over2$}}

\def\arr{\rightarrow}
\def\normal{{\cal N}}
\def\sumin{\sum_{i=1}^n}

\def\RR{\mathord{I\kern-.3em R}}
\def\PP{\mathord{I\kern-.3em P}}
\def\NN{\mathord{I\kern-.3em N}}
\def\ZZ{\mathord{I\kern-.3em Z}} 
\def\Var{{\rm Var}}
\def\E{{\rm E}}
\def\d{{\rm d}}

\def\mtrix{\pmatrix} 
\def\midd{{\,|\,}}
\def\subsection{\medskip}

\font\bigbf=cmbx12

\font\csc=cmcsc10

 at 10truept 
\font\smallrm=cmr8

\def\today{\number\day \space \ifcase\month\or
January\or February\or March\or April\or May\or June\or 
July\or August\or September\or October\or November\or December\fi  
\space \number\year}


   
\def\ref#1{{\noindent\hangafter=1\hangindent=20pt
  #1\smallskip}}          

\def\quotationone{\smallrm Where there is a Will}
\def\quotationtwo{\smallrm There is a Won't}
\def\hskipdistanceleft{\hskip-3.5pt}
\def\hskipdistanceright{\hskip-2.0pt}
\footline={{
\ifodd\count0
        {\hskipdistanceleft\quotationone\phantom{\smallrm\today}
                \hfil{\rm\the\pageno}\hfil
         \phantom{\quotationone}{\smallrm\today}\hskipdistanceright}
        \else 
        {\hskipdistanceleft\quotationtwo\phantom{\today}
                \hfil{\rm\the\pageno}\hfil
         \phantom{\quotationtwo}{\smallrm\today}\hskipdistanceright}
        \fi}}

         
\def\cstok#1{\leavevmode\thinspace\hbox{\vrule\vtop{\vbox{\hrule\kern1pt
        \hbox{\vphantom{\tt/}\thinspace{\tt#1}\thinspace}}
        \kern1pt\hrule}\vrule}\thinspace} 
\def\square{\cstok{\phantom{$\cdot$}}} 


\def\fermat#1{\setbox0=\vtop{\hsize4.00pc
        \smallrm\raggedright\noindent\baselineskip9pt
        \rightskip=0.5pc plus 1.5pc #1}\leavevmode
        \vadjust{\dimen0=\dp0
        \kern-\ht0\hbox{\kern-4.00pc\box0}\kern-\dimen0}}


\def\today{{August 1994}}

\def\quotationone{\smallrm Local Bayesian Regression}
\def\quotationtwo{\smallrm Nils Lid Hjort} 

\centerline{\bigbf Local Bayesian Regression} 

\medskip
\centerline{\bigbf Nils Lid Hjort, University of Oslo}

\medskip
{{\smallskip\narrower\noindent\baselineskip11pt
{\csc Abstract.}  
This paper develops a class of Bayesian non- and semiparametric 
methods for estimating regression curves and surfaces. 
The main idea is to model the regression as locally linear,
and then place suitable local priors on the local parameters.
The method requires the posterior distribution of the local 
parameters given local data, and this is found 
via a suitably defined local likelihood function.
When the width of the local data window is large 
the methods reduce to familiar fully parametric Bayesian methods, 
and when the width is small the estimators are essentially nonparametric.  
When noninformative reference priors are used 
the resulting estimators coincide with recently developed 
well-performing local weighted least squares methods 
for nonparametric regression. 

Each local prior distribution needs in general 
a centre parameter and a variance parameter. 
Of particular interest are versions of the scheme that are 
more or less automatic and objective in the sense that 
they do not require subjective specifications of prior parameters. 
We therefore develop empirical Bayes methods to obtain 
the variance parameter and a hierarchical Bayes method 
to account for uncertainty in the choice of centre parameter. 
There are several possible versions of the general programme,
and a number of its specialisations are discussed. 
Some of these are shown to be capable of outperforming
standard nonparametric regression methods, 
particularly in situations with several covariates. 

\smallskip\noindent
{\csc Key words:} \sl
Bayesian regression; 
empirical Bayes; 
hierarchical Bayes; 
kernel smoothing; 
local likelihood; 
locally linear models; 
Poisson regression; 
semiparametric estimation; 
Stein-type estimators 
\smallskip}} 

\bigskip
{\bf 1. Introduction and summary.}
Suppose data pairs $(x_1,y_1),\ldots,(x_n,y_n)$ are available 
and that the regression of $y$ on $x$ is needed, 
in a situation where there is, a priori, 
no acceptable simple parametric form for this. 
We take the regression curve to be the conditional mean function 
$m(x)=\E(y\midd x)$, and choose to work in the slightly more specific model 
where the $y_i$s are seen as regression curve 
plus i.i.d.~zero mean residuals. 
In other words, given $x_1,\ldots,x_n$, we have  
$$y_i=m(x_i)+\eps_i,
\quad i=1,\ldots,n, 
\quad {\rm where\ }\E\,\eps_i=0 {\rm\ and\ }\Var\,\eps_i=\sigma^2. \eqno(1.1)$$
This paper is about non- and semiparametric Bayesian approaches 
towards estimating the $m(\cdot)$ function. 

\subsection
{\csc 1.1. Two standard estimators.}  
Before embarking on the Bayesian journey we describe 
two standard frequentist estimation methods in some detail, 
since these will show up as important ingredients 
of some of the Bayes solutions to come. 
The first method uses 
$$\tilda m(x)
=\hbox{the $a$ that minimises\ }\sumin(y_i-a)^2w_i(x), \eqno(1.2)$$ 
that is, $\tilda m(x)=\sumin w_i(x)y_i/\sumin w_i(x)$, 
for a suitable set of {\it local weights} $w_i(x)$. 
These are to attach most weight to pairs $(x_i,y_i)$ 
where $x_i$ is close to $x$ and little or zero weight 
to pairs where $x_i$ is some distance away from $x$.  
A simple way of defining such weights is via a kernel function
$K(\cdot)$, some chosen probability density function.
Typical kernels will be unimodal and symmetric  
and at least continuous at zero. 
To fix ideas we also take $K$ to have bounded support, 
which we may scale so as to be $[-\half,\half]$.  
We choose to define 
$$w_i(x)=\bar K(h^{-1}(x_i-x)), 
        \quad {\rm where\ }\bar K(z)=K(z)/K(0). \eqno(1.3)$$
Of course scale factors do not matter in (1.2),  
but it is helpful both for interpretation and 
for later developments to scale the weights in this way.
We think of $w_i(x)$ as a measure of influence 
for data pair $(x_i,y_i)$ when estimation is 
carried out at location $x$, and with scaling as per (1.3),  
$w_i(x)$ is close to 1 for pairs where $x_i$ is near $x$
and equal to 0 for pairs where $|x_i-x|>\half h$. 
The local weighted mean estimator (1.2) is the 
Nadaraya--Watson (NW) estimator, see for example Scott (1992, Ch.~8) 
for another derivation and for further discussion. 

Method (1.2) fits a local constant to the local $(x_i,y_i)$ pairs.
The second standard method is the natural extension 
which fits a local regression line to the local data, that is, 
$$\tilda m(x)=\tilda a, 
\quad {\rm where\ }(\tilda a,\tilda b){\rm\ minimise\ }
\sum_{|x_i-x|\le h/2}\{y_i-a-b(x_i-x)\}^2w_i(x). \eqno(1.4)$$
See Wand and Jones (1994, Ch.~xx) for performance properties 
and further discussion of this local linear (LL) regression estimator. 
Obviously the idea generalises further, 
to other local parametric forms for the regression curve, 
to other criterion functions to minimise,
and to other regression models. 

\subsection
{\csc 1.2. Local likelihood functions.} 
The basic Bayesian idea to be developed 
is to place local priors on the local parameters, 
and use posterior means as estimators. 
This requires a `local likelihood' function that expresses
the information content of the local data. 
Suppose in general terms that there is some parametric 
model $f(y_i\midd x_i,\beta,\sigma)$, typically in terms of 
regression parameters $\beta$ and a scale parameter $\sigma$,
that is trusted locally around a given $x$, that is, 
for $x_i\in N(x)=[x-\half h,x+\half h]$. 
The likelihood for these local data are then 
$\prod_{x_i\in N(x)}f(y_i\midd x_i,\beta,\sigma)$.
This can also be written
$$L_n(x,\beta,\sigma)=\prod_{x_i\in N(x)} 
        f(y_i\midd x_i,\beta,\sigma)^{w_i(x)}, \eqno(1.5)$$
with weights as in (1.3) for the uniform kernel on $[-\half,\half]$.
We will also use (1.5) for more general kernel functions, 
requiring only that $\bar K$ is continuous at zero
with `correct level' $\bar K(0)=1$, 
and call $L_n(x,\beta,\sigma)$ the local kernel smoothed likelihood at $x$. 
The argument is that it is a natural smoothing generalisation of 
the bona fide one that uses uniform weights, 
and that the local parametric model employed is 
sometimes to be trusted less a little distance away from $x$ 
than close to $x$. 
Further motivation is provided by the fact that 
the resulting maximum local likelihood estimators 
sometimes have better statistical behaviour 
for non-uniform choices of kernel function;
the standard local likelihood with uniform weights is for
example not continuous in $x$. 
See further comments in Sections 8.3--8.4 below. 
Also note that when $h$ grows large all weights become equal to 1,
and we are back in familiar fully parametric territory. 

Observe that when the local model is normal with constant variance, 
then the `local constant mean' model gives a local likelihood 
with maximum equal to the NW estimator (1.2), 
and the `local linear mean' model correspondingly gives the 
LL estimator (1.4) interpretation as maximum local likelihood estimator. 
This is discussed more fully in Sections 2 and 3. 

\subsection
{\csc 1.3. General Bayesian construction.} 
A `locally parametric Bayesian regression programme' 
can be outlined quite broadly, but we will presently do so 
in terms of a local vehicle model of the form $f(y\midd x,\beta,\sigma)$. 
It comprises several different and partly related steps. 

\smallskip
\item{(a)} 
Come up with a complete `start curve' function $m_0(x)$,
and give a prior for the scale parameter $\sigma$. 
This start curve is thought of as any plausible candidate 
or `prior estimate' for the regression function $m(x)$,
and would also be called the `prior guess function' 
in Bayesian parlance.    

\item{(b)} 
For each given $x$, place a prior on the local regression parameter
$\beta=\beta_x$, typically a normal with a centre parameter determined 
by the start curve function and a covariance matrix of the form 
$\sigma^2W_{0,x}^{-1}$. 

\item{(c)} 
Do the basic Bayesian calculation and obtain the Bayes estimate
$\hatt m(x)$ as the local posterior mean, that is,
the mean in the distribution of $\E(y\midd x,\beta,\sigma)$ 
given local data. Further inference (finding credibility intervals and so on) 
can be carried out using the same distribution. 
 
\smallskip\noindent 
This is `so far, so good', and the problem is solved for the 
`ideal Bayesians' who can accurately specify the $m_0(\cdot)$ function
and the precision parameters $W_{0,x}$. 
Step (c) will indeed give estimators of interest. 
Not every practising statistician can come up with the required 
parameters of the local priors, however. 
We therefore add two more steps to the programme:

\smallskip
\item{(d)} 
Obtain estimates of local precision parameters $W_{0,x}$ 
using empirical Bayes calculations, and still using the 
start curve function $m_0(x)$. 

\item{(e)} 
To account for uncertainty in specifying the start curve,
and to obtain an estimate thereof,  
use a hierarchical Bayesian approach, 
by having a background prior on this curve. 

\smallskip\noindent 
There will be several possible versions of (d) and (e) here.
We shall be concerned with versions of (e) 
that are in terms of a parametric start curve function $m_0(x,\xi)$,
with a first-stage prior on the $\xi$ parameter. 
We shall also be primarily interested in versions of (d) and (e)
that are reasonably `automatic' and objective in the sense
that they do not require subjective specification of prior 
parameters. That is, a flat prior will typically be used for the
parameters present in $m_0(x,\xi)$, and the empirical Bayes 
methods in (d) will typically be based only on the unconditional
distribution for collections of local data, and not, for example,
on further priors for the parameters in question. 
Modifications are available under strong prior opinions, though.  

\subsection
{\csc 1.4. Other work.} 
Local regression methods go back at least to  
Stone (1977) and Cleveland (1979).
See Fan and Gijbels (1992), 
Hastie and Loader (1993), Ruppert and Wand (1994) 
and Wand and Jones (1994) for recent developments. 
Local likelihood methods were first explicitly introduced 
by Tibshirani, see Tibshirani and Hastie (1987) 
and the brief discussion in Hastie and Tibshirani (1990, Ch.~6). 
Fully parametric Bayesian regression methods, 
corresponding in the present context to a large window width $h$, 
are well known, see for example Box and Tiao (1973). 
Bayesian versions of local regression methods 
do not seem to have been considered before. 
Similarly spirited Bayesian locally parametric estimation methods
for hazard rates and probability densities 
have however been proposed and discussed in Hjort (1994b),
using suitable local likelihood constructions for such,
developed in respectively Hjort (1994a) and Hjort and Jones (1994).  
The main Bayesian-like nonparametric regression method
is that of splines, see Silverman (1985), Wahba (1990) 
and the discussion in Hastie and Tibshirani (1990, Ch.~3).
Besides splines there does not seem to have been much work 
done in semi- and nonparametric Bayesian regression at all; 
a recent review paper on general Bayesian nonparametric methods 
by Ferguson, Phadia and Tiwari (1992) barely touches regression. 
Some methods have recently been developed using mixtures of 
Dirichlet processes, see 
Erkanli, M\"uller and West (1994) and 
West, M\"uller and Escobar (1994). 

\subsection
{\csc 1.5. The present paper.} 
Section 2 goes through the Bayesian regression programme 
for the `local constant' model, which is the most transparent case.  
It gives specific proposals for interpolating between 
a given start curve and the NW estimator,
with weighting schemes that come from empirical Bayes considerations
and that have goodness of fit interpretations.   
In the end the start curve is averaged over with respect to
its posterior distribution. 
Section 3 similarly treats the `local linear' model,
where somewhat more technical calculations are called for. 
Bayesian and empirical Bayesian generalisations of the LL estimator 
(1.4) are obtained. 
To illustrate the various ingredients in the complete estimation
programme a case of a linearly structured start curve 
is studied in Section 4. 
Other specialisations of the scheme are considered in Section 5. 
One particular version of interest models the regression curve
locally as being of the form a start curve, 
say with globally estimated parameters, times a local correction factor,
and produces in the end estimators that resemble Bayesian relatives
of a frequentist method recently proposed in Hjort and Glad (1994).  
The Bayesian methods might be especially fruitful in situations
with several covariates, since many of the standard methods 
based on local smoothing have severe difficulties then. 
This is briefly discussed in Section 6. 

To show that the general apparatus also works well in 
regression models other than the traditional one, 
we consider Poisson regression in some detail in Section 7,
and give brief pointers to other types of applications.
Section 8 presents some additional results and remarks.
Matters dwelt with include fine-tuning of parameters, 
parallels to Stein-type shrinkage estimators,
and discussion of the kernel smoothed local likelihood approach. 
Some of the comments suggest problems for further research.
Finally conclusions are offered in Section 9. 

\bigskip
{\bf 2. Inference for the `local level' model.}
Let the local model be the normal with constant variance  
and local mean function $m(t)=a$ for $t\in N(x)$. 
The local kernel smoothed likelihood becomes 
$$\eqalign{L_n(x,a,\sigma)
&=\prod_{x_i\in N(x)}
\Bigl[{1\over \sigma}\exp\Bigl\{-\half{1\over \sigma^2}
        (y_i-a)^2\Bigr\}\Bigr]^{w_i(x)} \cr 
&=\Bigl({1\over \sigma}\Bigr)^{s_0(x)}
        \exp\Bigl\{-\half{1\over \sigma^2}Q(x,a)\Bigr\} \cr}$$
(ignoring constant factors), where $s_0(x)=\sum_{N(x)}w_i(x)$ and 
$Q(x,a)=\sum_{N(x)}(y_i-a)^2w_i(x)$. 
Note that the maximum local likelihood estimator 
is equal to the NW estimator of (1.2), $\tilda m(x)=\tilda m_{\rm NW}(x)$. 

\subsection
{\csc 2.1. Basic local Bayesian calculation.}
As the local prior for $a$ at $x$ we use a normal 
with mean $m_0(x)$ and variance $\sigma^2/w_0$. 
The precision parameter $w_0$ will be allowed to vary with $x$, 
see the following subsections, but at the moment $x$ is fixed. 
Start out rewriting 
$$\eqalign{Q(x,a)
&=\sum_{N(x)}\{y_i-\tilda m(x)\}^2w_i(x)
        +\sum_{N(x)}\{a-\tilda m(x)\}^2w_i(x) \cr
&=Q_0(x)+s_0(x)\{a-\tilda m(x)\}^2. \cr} \eqno(2.1)$$
Using this it is not difficult to derive that $a$ 
given the local data, and conditional on $\sigma$, 
is another normal, centred at 
$$\eqalign{\hatt m(x)
&=\E\{a\midd{\rm local\ data},\sigma\}
        ={w_0m_0(x)+s_0(x)\tilda m(x)\over w_0+s_0(x)} \cr
&=\rho(x)m_0(x)+\{1-\rho(x)\}\tilda m(x). \cr}\eqno(2.2)$$
This is the Bayes estimator (since it does not depend on $\sigma$),  
as per Step (c) in the general scheme described in Section 1.3. 

While the local posterior mean (2.2) is the essential ingredient,
as far as computation of the Bayes estimate is concerned, 
we also go to the trouble of noting the following expressions 
for the simultaneous density of $a=a_x$ and local data,
conditional on $\sigma$: 
$$\eqalign{
&{w_0^{1/2}\over \sigma}\Bigl({1\over \sigma}\Bigr)^{s_0(x)}
\exp\Bigl\{-\half{1\over \sigma^2}
\Bigl[w_0\{a-m_0(x)\}^2+s_0(x)\{a-\tilda m(x)\}^2+Q_0(x)\Bigr]\Bigr\} \cr
=&{\{w_0+s_0(x)\}^{1/2}\over \sigma}
\exp\Bigl[-\half{1\over \sigma^2}\{w_0+s_0(x)\}\{a-\hatt m(x)\}^2\Bigr] \cr
 &{w_0^{1/2}\over \{w_0+s_0(x)\}^{1/2}}
\Bigl({1\over \sigma}\Bigr)^{s_0(x)}
\exp\Bigl\{-\half{1\over \sigma^2}\Bigl[Q_0(x)+{w_0s_0(x)\over w_0+s_0(x)}
        \{\tilda m(x)-m_0(x)\}^2\Bigr]\Bigr\}. \cr}$$
This will be useful later in connection with estimation of $\sigma$
and specification of $w_0$.   
And, in particular, a fuller description of the local posterior is 
$$m(x)\midd{\rm local\ data,\sigma}\sim\normal\bigl[\hatt m(x),
        \sigma^2/\{w_0+s_0(x)\}\bigr]. \eqno(2.3)$$
With a Gamma prior for $1/\sigma^2$ this also leads to 
a suitable $t$ distribution for $a$ given local data; see Section 8.2. 

The Bayes estimator (2.2) is a convex combination 
of start curve and the NW estimator.
It pushes the NW estimator towards the start curve  
with a strength determined by the prior precision parameter $w_0$.
Note that the data strength parameter $s_0(x)$ can be 
expressed as $nhf_n(x)/K(0)$, 
where $f_n(x)=n^{-1}\sumin h^{-1}K(h^{-1}(x_i-x))$ is 
the classical kernel estimator of the density $f$ for the $x$s. 
The NW estimator corresponds to having 
$w_0$ close to or equal to zero, which means 
a flat and noninformative prior for the local level. 
Also note that when $h$ is large, 
then $\tilda m(x)$ is the mean of the $y_i$s and $s_0(x)=n$,
and the Bayes solution is as for the familiar fully parametric case.   

It is intuitively clear that the Bayes solution $\hatt m(x)$ 
has better precision than the standard estimator $\tilda m(x)$
as long as the true value $m(x)$ is not too far from 
the prior estimate value $m_0(x)$. 
A formal investigation of this can study their risk functions 
under squared error loss, that is, the two mean squared errors.
This is done in Section 8.1 below. 
For the present moment note that 
$$\E_{m(x)}\{\tilda m(x)-m(x)\}^2=\sigma^2t_0(x)/s_0(x)^2, 
\quad {\rm where\ }t_0(x)=\sum_{N(x)}w_i(x)^2. $$ 
This last quantity can be written $nhR_K/K(0)^2$ times a kernel
estimate of $f$, where $R_K=\int K(z)^2\,\d z$, 
so the mean squared error is approximately $R_K\sigma^2/\{nhf(x)\}$ 
(under the constant local mean model). 
The point is that the standard estimator is perhaps acceptable 
in regions of high $x_i$-density, but quite variable 
in regions of low $x_i$-density. 
This indicates that the Bayes estimate,
which has expected risk $\sigma^2/\{w_0+s_0(x)\}$,  
is likely to make a significant improvement in situations 
where $f$ is small, where $w_0$ is not small, 
and where the start curve value $m_0$ is not far off. 
In the following methods will be given that make both
$w_0$ and $m_0$ dictated by data.  

\subsection
{\csc 2.2. Estimating $\sigma$ and prior precision parameters.} 
The Bayes solution (2.2) needs specification of 
the prior strength parameter $w_0$. 
In some situations one could perceivably specify this number based
on previous data sets or other prior considerations,
as in purist Bayes analysis. It is of considerable interest to 
develop more automatic and data-dictated methods, however.
Information about the scale parameter $\sigma$ is also needed, 
in the form of an estimate or a posterior density,  
in order to carry out further inference about $m(x)$, see (2.3).  


The empirical Bayes idea is to infer parameters of the prior 
from the unconditional distribution of data. 
Given the local constant level $a$, 
the NW estimator has mean $a$ and variance $\sigma^2t_0(x)/s_0(x)^2$.
It follows that its unconditional mean is $m_0(x)$ and
unconditional variance $\sigma^2\{t_0(x)/s_0(x)^2+1/w_{0,x}\}$, that is,
$$P_0(x)=s_0(x)\{\tilda m(x)-m_0(x)\}^2
\quad {\rm has\ }
\E\{P_0(x)\midd\sigma\}=\sigma^2\Bigl\{
        {t_0(x)\over s_0(x)}+{s_0(x)\over w_{0,x}}\Bigr\}, \eqno(2.4)$$
writing now $w_{0,x}$ for the precision parameter at $x$. 
With an estimate of $\sigma$ this can be used to assign values 
for different $w_{0,x}$s. 
We shall arrive at versions of this scheme below. 

The distribution of local data ${\cal D}(x)$ alone, 
given $\sigma$ and $w_{0,x}$, is obtained by integrating out $a$ 
from the expression that led to (2.3). The result is 
$$\rho(x)^{1/2}\sigma^{-s_0(x)}
\exp[-(2\sigma^2)^{-1}\{Q_0(x)+\rho(x)P_0(x)\}], $$
in terms of the $\rho(x)=w_{0,x}/\{w_{0,x}+s_0(x)\}$ parameter. 
This can be maximised over possible values 
of $\sigma$ and $w_{0,x}$, with result 
$$\tilda\sigma_x^2={Q_0(x)\over s_0(x)-1}
\quad {\rm and} \quad 
\tilda\rho(x)={\tilda\sigma_x^2\over P_0(x)}. $$
The $\tilda\rho(x)$ is set equal to 1 if $\tilda m(x)$ 
is so close to $m_0(x)$ that the ratio exceeds 1. 
This $\sigma$ estimate is local to $x$ and of some 
separate interest, for example for checking of the 
constant variance assumption. 
A better estimate emerges by combining information over many
neighbourhoods. Divide the interval in which data fall into 
say $k$ such cells $N(x)$,
say with midpoints $x_{0,1}<\cdots<x_{0,k}$ and lengths $h_1,\ldots,h_k$. 
Let $I$ be the collection of these midpoints, so that 
$${\rm full\ data\ interval\ }=\cup_{x\in I}N(x)
\quad {\rm and} \quad 
  \{x_1,\ldots,x_n\}=\cup_{x\in I}{\cal D}(x). $$
Assume now that the local levels $a_1,\ldots,a_k$ 
at $x_{0,1},\ldots,x_{0,k}$ are taken to be independent in
their joint prior distribution 
(and see Section 5.4 for an alternative). 
This leads to a combined likelihood of the form 
$$\prod_{x\in I}\Bigl\{\rho(x)^{1/2}\sigma^{-s_0(x)}
\exp[-(2\sigma^2)^{-1}\{Q_0(x)+\rho(x)P_0(x)\}]\Bigr\}, $$
conditional on $\sigma$ and the $w_{0,x}$s 
at the different midpoints locations. 
Some analysis reveals that this is maximised for 
$$\tilda\sigma^2={\sum_{x\in I}Q_0(x)\over \sum_{x\in I}\{s_0(x)-1\}}
\quad {\rm and} \quad 
\tilda\rho(x)={\tilda\sigma^2\over s_0(x)}
        {1\over \{\tilda m(x)-m_0(x)\}^2}
        ={\tilda\sigma^2\over P_0(x)}. \eqno(2.5)$$
Notice that this agrees very well with (2.4), 
especially when uniform weights are used, in which case $t_0(x)/s_0(x)=1$.
Note that the $\sigma$ estimate is independent of start curve $m_0$,
whereas the $\rho(x)$ estimate quite naturally measures 
fit of data to the start curve, locally at $x$. 
Again the $\tilda\rho(x)$ is truncated at 1; 
if $s_0(x)^{1/2}|\tilda m(x)-m_0(x)|<\tilda\sigma$,
then the prior model fits excellently at $x$, 
and one uses $\tilda\rho(x)=1$, or $\hatt w_{0,x}=\infty$, in (2.5). 
If $P_0(x)^{1/2}>\tilda\sigma$, 
then the Bayes-empirical-Bayes estimate becomes 
$$\hatt m(x)={\tilda\sigma^2\over P_0(x)}m_0(x)
+\Bigl\{1-{\tilda\sigma^2\over P_0(x)}\Bigr\}\tilda m(x)
=\tilda m(x)-{\tilda\sigma^2\over s_0(x)}
        {1\over \tilda m(x)-m_0(x)}. \eqno(2.6)$$

Often the prior strength parameter $w_{0,x}$ would naturally be 
considered to be a smoothly changing function with $x$, and then 
the estimate implicitly given above will be too rugged. 
One may use various post-smoothing devices for $w_{0,x}$ or 
its relative $\rho(x)$.
The local goodness of fit statistics $P_0(x)$ 
have been scaled so as to have approximately the same variance,
which invites using $\tilda\sigma^2$ divided by 
an average of these as a single measure of the overall faithfulness 
of observed data to the start curve used. 
This leads to an estimator of the form 
$$\eqalign{\hatt m(x)
&={k\tilda\sigma^2\over 
\sum_{x\in I} s_0(x)\{\tilda m(x)-m_0(x)\}^2}m_0(x) \cr
&\qquad\qquad 
+\Bigl[1-{k\tilda\sigma^2\over 
\sum_{x\in I} s_0(x)\{\tilda m(x)-m_0(x)\}^2}\Bigr]\tilda m(x), \cr} 
                \eqno(2.7)$$ 
where both weights are truncated to $[0,1]$ if necessary. 
The Steinean overtones already discernible above are now more audible;
see Section 8.1 below for further discussion. 

Of course non-uniform averaging is sometimes more appropriate 
when forming the denominator here. A satisfactory solution would 
in many situations be to model the prior variance $\sigma^2/w_{0,x}$ 
at $x$ in a parametric fashion.
The strategy is to estimate the necessary parameters
either by regressing $P_0(x)/\tilda\sigma^2$ against 
$t_0(x)/s_0(x)+s_0(x)/w_{0,x}$ at different $x$ positions, 
or by maximising the combined likelihood 
$$\prod_{x\in I}\Bigl\{{w_{0,x}\over w_{0,x}+s_0(x)}\Bigr\}^{1/2}
\Bigl({1\over \sigma}\Bigr)^{\sum_{x\in I}s_0(x)} 
\exp\Bigl[-\half{1\over \sigma^2}\Bigl\{\sum_{x\in I}Q_0(x)
+\sum_{x\in I}{w_{0,x}s_0(x)P_0(x)\over w_{0,x}+s_0(x)}\Bigr\}\Bigr] $$
with respect to the parameters present in $w_{0,x}$. 
In the end one uses (2.2) with inferred values for $w_{0,x}$
(or, if one prefers, inferred values for $\rho(x)$). 
For an example, suppose the prior variance at $x$ is modelled 
as $\sigma^2r(x)/w_0$ for a known function $r(x)$. 
Plugging in $w_{0,x}=w_0/r(x)$ in the combined likelihood 
leads to one feasible solution, but a simpler one is based on 
$P_0(x)/\tilda\sigma^2\simeq 1+s_0(x)r(x)/w_0$,
assuming uniform weights to be used.  
Since the $P_0(x)$s have the same variability it is natural to 
equate the clean average $\bar P_0/\tilda\sigma^2$ to $1+c/w_0$ 
to define the empirical Bayes estimate $\hatt w_0$,
where $\bar P_0=k^{-1}\sum_{x\in I}P_0(x)$ 
and $c=k^{-1}\sum_{x\in I}s_0(x)r(x)$. This leads to 
$$\hatt m(x)={m_0(x)\over 1+s_0(x)r(x)c^{-1}\{\bar P_0/\tilda\sigma^2-1\}}
+{s_0(x)r(x)c^{-1}\{\bar P_0/\tilda\sigma^2-1\}\tilda m(x)
        \over 1+s_0(x)r(x)c^{-1}\{\bar P_0/\tilda\sigma^2-1\}}. \eqno(2.8)$$
This is quite similar to (2.7). 
If in particular the prior variance at $x$ is seen 
as approximately inversely proportional to the density $f(x)$ at $x$, 
then $r(x)s_0(x)$ is approximately constant, and (2.8) reduces to (2.7). 
As earlier the weights of this Stein-type estimator are truncated to $[0,1]$. 

The empirical Bayesian attitude in this subsection 
has perhaps been more `classical' than `pure Bayesian'; 
unconditional likelihoods have been maximised 
rather than used in connection with additional priors. 
Maximising the likelihood is equivalent to using the Bayes
solution with a flat prior for the parameters,
under a sharp 0--1 loss function, 
so the procedures suggested above for getting hold of $w_{0,x}$s 
can be viewed as Bayesian but, consciously, 
with no additional prior information about $\sigma$ or the $w_{0,x}$s. 
Bayesian analysis with a Gamma prior for $1/\sigma^2$ 
is technically convenient, and it is reassuring to see 
that the noninformative prior version of this 
leads to exactly the same estimates as in (2.5).
More generally, maximising the derived likelihood for data alone, 
with respect to parameters present in the $w_{0,x}$s, 
gives the same results as when maximising over the 
joint likelihood that includes $\sigma$. 
Yet other ways of estimating $\sigma$ and $w_{0,x}$s 
are briefly discussed in Section 8.2. 

\subsection
{\csc 2.3. Prior uncertainty around the start curve.}
The estimator that has so far been developed 
is the Bayes estimator (2.2) with inserted empirical Bayes estimate,  
say $\hatt w_{0,x}$, for prior precision.   
The start curve function $m_0(x)$ enters both (2.2) and the 
$\hatt w_{0,x}$ operation crucially. 
There is usually uncertainty around the choice of the $m_0$ curve. 
A two-stage prior framework is to view $m_0(\cdot)$ 
as the result of some background prior process. 
The final estimator is then, in principle, to average 
the estimator just described over the posterior distribution 
for $m_0(\cdot)$ given all data. 

Suppose that $m_0(\cdot)$ is modelled parametrically, say 
$m_0(x,\xi)$, with a background prior $\pi_0(\xi)$ for $\xi$. 
The regression curve estimator is 
$$\hatt m(x,\xi)={\hatt w_{0,x}(\xi)m_0(x,\xi)+s_0(x)\tilda m(x)
        \over \hatt w_{0,x}(\xi)+s_0(x)}, $$
conditional on $\xi$. The final estimator is accordingly 
$$\eqalign{\hatt m(x)
&=\E\{\hatt m(x,\xi)\midd{\rm all\ data}\} \cr
&=\int{\hatt w_{0,x}(\xi)m_0(x,\xi)+s_0(x)\tilda m(x)
\over \hatt w_{0,x}(\xi)+s_0(x)}
        \pi_0(\xi\midd{\rm all\ data})\,\d\xi. \cr}\eqno(2.9)$$
This would have to be computed through numerical integration
or simulation. 
In situations where a sufficient amount of good data gives 
a reasonably concentrated posterior density 
$\pi_0(\xi\midd{\rm all\ data})$,
say around the Bayes estimate $\hatt\xi$, 
a simple approximation is to plug this in, and use 
$$\hatt m(x,\hatt\xi)
={\hatt w_{0,x}(\hatt\xi)m_0(x,\hatt\xi)+s_0(x)\tilda m(x)
        \over \hatt w_{0,x}(\hatt\xi)+s_0(x)}. $$
Further approximations could also be contemplated,
for example using a quadratic approximation of the integrand 
around the Bayes estimate for $\xi$. 

We need to comment on what is meant by 
the density of $\xi$ given all data here. 
One could base this on the parametric likelihood 
$\sigma^{-n}\exp[-(2\sigma^2)^{-1}\sumin\{y_i-m_0(x_i,\xi)\}^2]$,
but this is not entirely satisfactory, since the point 
in the end is to be nonparametric; 
the $m_0(x,\xi)$ model is only meant as an initial description. 
It is better, therefore, to view $\pi_0(\cdot)$ as a prior for 
the `least false' parameter vector $\xi_0$ that gives best parametric 
approximation to the true $m(x)$ curve, in the sense that it
minimises the long-term version of $n^{-1}\sumin\{m(x_i)-m_0(x_i,\xi)\}^2$.
The approximate distribution of the maximum likelihood estimator 
$\tilda\xi$, say $L_n(\tilda\xi\midd\xi_0)$, can often be worked out, 
even outside the conditions of the parametric model, 
see Hjort and Pollard (1994, Sections 3 and 4).
This suggests employing the distribution of $\xi$ given its maximum likelihood
estimate $\tilda\xi$ in (2.9), that is, using 
$\pi_0(\xi)L_n(\tilda\xi\midd\xi)/\int \pi_0(\xi)L_n(\tilda\xi\midd\xi)\,\d\xi$
for $\pi_0(\xi\midd{\rm all\ data})$. 

For a specific example that illustrates these calculations, 
suppose that $m_0(x,\xi)$ is modelled as 
$\xi_1+\xi_2g_2(x)+\cdots+\xi_pg_p(x)$,
in terms of given basis functions $g_2,\ldots,g_p$.
The explicit maximum likelihood estimator is 
$\tilda\xi=(\sumin z_iz_i')^{-1}\sumin z_iy_i$, 
where $z_i=z(x_i)$ is the $p$-vector $(1,g_2(x_i),\ldots,g_p(x_i))'$.
Under mild technical assumptions the estimator is 
consistent for the least false parameter vector $\xi_0$ described above; 
indeed $\xi_0$ can be expressed as $\{\E z(x_i)z(x_i)'\}^{-1}\E z(x_i)y_i$.  
Results from Hjort and Pollard (1994, Section 3) 
imply furthermore that it is approximately a multinormal, 
centred at $\xi_0$ and with a covariance matrix $\tilda V/n$, where 
$$\tilda V=\Bigl(n^{-1}\sumin z_iz_i'\Bigr)^{-1}
\Bigl[n^{-1}\sumin \{y_i-m_0(x_i,\tilda\xi)\}^2z_iz_i'\Bigr]
\Bigl(n^{-1}\sumin z_iz_i'\Bigr)^{-1}. \eqno(2.10)$$
The present use of this is that if the parameter vector 
$\xi$ is given a suitable prior, 
for example a uniform reference prior,
and viewed necessarily as a prior for the least false $\xi_0$, 
then the posterior distribution is approximately a multinormal,
centred at the estimate $\tilda\xi$ and with the same
covariance matrix $\tilda V/n$ as above. 
This follows from general arguments and results provided in 
Hjort and Pollard (1994, Section 4). 
The approximation is best when the prior 
is the noninformative uniform one, but is valid for any 
fixed prior as long as its covariance matrix is 
large compared to $\tilda V/n$. Modifications are available
to account for strong prior opinions about $\xi$. 

The final Bayes regression estimator (2.9) can now be computed as follows.
Draw say 100 values of $\xi$ from the $\normal_p\{\tilda\xi,\tilda V/n\}$
distribution. This gives 100 start curves $m_0(x,\xi)$
that are all likely given the full data information. 
For each of the 100 likely start curves 
the algorithm in question is used to compute the $\hatt m(x,\xi)$ curve. 
In the end these are averaged.
See Section 4 below for illustrations. 

The discussion above used $L_n(\tilda\xi\midd\xi)$ 
to get to $\pi_0(\xi\midd{\rm all\ data})$. 
An alternative route worthy of at least a brief discussion is to 
use the marginal distribution of the combined local data sets ${\cal D}(x)$
as the likelihood for the $\xi$ parameters present in $m_0(\cdot)$. 
The relevant part of this likelihood becomes 
$$\exp\Bigl\{-\half{1\over \sigma^2}\Bigl[\sum_{x\in I}
{w_{0,x}s_0(x)^2\over w_{0,x}+s_0(x)}
        \{\tilda m(x)-\xi'z(x)\}^2\Bigr]\Bigr\}. $$ 
Some calculations transfer this to a multinormal likelihood, 
with mean parameter 
$B^{-1}\sum_{x\in I}\rho(x)\allowbreak s_0(x)z(x)\tilda m(x)$ 
and with covariance matrix $\sigma^2B^{-1}$, where 
$B=\sum_{x\in I}\rho(x)s_0(x)z(x)z(x)'$. 
One may now show that the posterior density of $\xi$ derived 
using this likelihood is approximately the same as the one 
used above, {\it if} the $m_0(x,\xi)$ model is approximately correct. 
We stick to the first method, mainly since we aim at 
valid inference outside the conditions of any narrow parametric model. 
Another nice facet of the first method, 
in view of the calculations just discussed, 
is that it does not need any extra binning of data. 

\smallskip
{\csc Remark.} The development above illustrates the crucial
Step (e) of the Bayesian programme, as outlined in Section 1.3, 
with a prior for a parametrised start curve. This is carried 
further in Section 4 below. One might also use a nonparametric
prior process for the start curve $m_0(\cdot)$, for example 
in the particular Gau\ss ian form that leads to splines;
see Hastie and Tibshirani (1990, Ch.~3) for some discussion.
Given data the curve is Gau\ss ian with specified parameters 
and it is possible to simulate from this posterior. This makes 
it possible to obtain the final Bayes regression curve 
in analogy to (2.9). This nonparametric posterior distribution 
would have larger variance than in the parametric case, however,
and the whole procedure ends up being a nonparametric attempt 
at correcting a nonparametric construction. In many cases 
such a method would probably not accomplish very much, 
and we view the benefits of working with a parametrised start 
curve, where a nonparametric correction is performed on 
a parametric initial construction, as more promising. \square 

\bigskip
{\bf 3. Inference for the local linear model.}
The local model worked with in the present section is 
again the normal with constant variance, 
but this time with a locally linear regression, say 
$m(t)=a+b(t-x)$ for $t\in N(x)$. In this section 
$(\tilda m(x),\tilda b(x))=(\tilda m_{\rm LL}(x),\tilda b_{\rm LL}(x))$ 
is the $(\tilda a,\tilda b)$ computed from the LL method (1.4) at $x$. 

\subsection
{\csc 3.1. Prerequisites.} 
The story to evolve is quite similar to that of Section 2, 
but with somewhat more involved algebraic calculations. 
We start with the local kernel smoothed likelihood, which is 
$\sigma^{-s_0(x)}\exp\{-(2\sigma^2)^{-1}Q(x,a,b)\}$, where 
$$Q(x,a,b)=\sum_{N(x)}\{y_i-a-b(x_i-x)\}^2w_i(x). $$
We need to be more technically specific about 
the maximum local likelihood estimators here, 
which are the ones already mentioned in (1.4).
Introduce 
$$S(x)=\mtrix{s_0(x) &s_1(x) \cr s_1(x) &s_2(x)}, $$
where $s_0(x)$, $s_1(x)$ and $s_2(x)$ are the local sums 
of respectively $w_i(x)$, $w_i(x)(x_i-x)$ and $w_i(x)(x_i-x)^2$. Then 
$$\mtrix{\tilda a \cr \tilda b}
=\mtrix{s_0(x) &s_1(x) \cr s_1(x) &s_2(x)}^{-1}
\mtrix{\sum_{N(x)}w_i(x)y_i \cr \sum_{N(x)}w_i(x)(x_i-x)y_i \cr}, $$
and the LL regression estimate itself is 
$$\tilda m(x)=\tilda a
=\{s_0(x)s_2(x)-s_1(x)^2\}^{-1}\sum_{N(x)}
w_i(x)\{s_2(x)-s_1(x)(x_i-x)\}y_i. $$
Some comments on the relative sizes of $s_0$, $s_1$, $s_2$ 
here are given in Section 8.6. 

\subsection
{\csc 3.2. Local posterior calculation.}
Start out with a normal prior for the local $(a,b)$ 
with mean $(a_0,b_0)$ and covariance matrix $\sigma^2W_0^{-1}$. 
Here $a_0=m_0(x)$ and $b_0=m_0'(x)$, 
and again $W_0$ will be allowed to depend on $x$ later on. 
Rewrite the exponent in the local likelihood as 
$$Q(x,a,b)=Q_0(x)+\mtrix{a-\tilda a \cr b-\tilda b}'S(x)
        \mtrix{a-\tilda a \cr b-\tilda b}, $$
where $Q_0(x)=\sum_{N(x)}\{y_i-\tilda a-\tilda b(x_i-x)\}^2w_i(x)$. 
Calculations analogous to those that led to (2.4) give that 
$(a,b)$ and local data ${\cal D}(x)$ have simultaneous distribution 
proportional to 
$$\eqalign{
&{|W_0+S(x)|^{1/2}\over \sigma^2}
\exp\Bigl[-\half{1\over \sigma^2}
\mtrix{a-\hatt a \cr b-\hatt b}'\{W_0+S(x)\}
\mtrix{a-\hatt a \cr b-\hatt b}\Bigr] \cr
\times&{|W_0|^{1/2}\over |W_0+S(x)|^{1/2}}
\Bigl({1\over \sigma}\Bigr)^{s_0(x)}
\exp\Bigl\{-\half{1\over \sigma^2}\Bigl[Q_0(x) \cr
&\qquad\qquad\qquad\qquad
+\mtrix{\tilda a-a_0 \cr \tilda b-b_0}'
\{W_0^{-1}+S(x)^{-1}\}^{-1}
\mtrix{\tilda a-a_0 \cr \tilda b-b_0}\Bigr]\Bigr\}, \cr}$$
where 
$$\eqalign{\mtrix{\hatt a \cr \hatt b}
&=\{W_0+S(x)\}^{-1}\Bigl\{W_0\mtrix{a_0 \cr b_0}
        +S(x)\mtrix{\tilda a \cr \tilda b}\Bigr\} \cr
&=\{I+W_0^{-1}S(x)\}^{-1}\mtrix{a_0 \cr b_0}
        +\{I+W_0^{-1}S(x)\}^{-1}W_0^{-1}S(x)
        \mtrix{\tilda a \cr \tilda b}. \cr} \eqno(3.1)$$
In particular,
$$\mtrix{a \cr b \cr}\,\Big|\,{\rm local\ data},\sigma
        \sim\normal_2\{\mtrix{\hatt a \cr \hatt b \cr},
        \sigma^2\{W_0+S(x)\}^{-1}\}, \eqno(3.2)$$
and the Bayes solution is $\hatt m(x)=\hatt a$. 

This is the appropriate generalisation of (2.2) and (2.3),
and the remarks made about the structure and characteristics 
of the Bayes solution, at the end of Section 2.1, 
are valid in the present case too, with suitable modifications. 
Note in particular that if $W_0$ tends to zero, 
signifying a noninformative prior for the local $(a,b)$,
then the Bayes solution is simply the LL estimator.
The second special case to note is that if $h$ is large, 
then we are again back in full parametric analysis in the 
linear normal model. 

\subsection
{\csc 3.3. Estimating $\sigma$ and local prior precision parameters.}
For a given start curve one needs to assign values to 
the prior precision matrices $W_0=W_{0,x}$.   
In analogy to (2.4), consider the matrix 
$$P_0(x)=\mtrix{\tilda m(x)-m_0(x) \cr \tilda b(x)-m_0'(x)}
\mtrix{\tilda m(x)-m_0(x) \cr \tilda b(x)-m_0'(x)}'S(x)
        =d(x)d(x)'S(x). \eqno(3.3)$$
Conditional on the local $(a,b)$, the $d(x)$ vector has mean 
$(a,b)$ and covariance matrix $S(x)^{-1}T(x)S(x)^{-1}$, where 
$$T(x)=\mtrix{\sum_{N(x)}w_i(x)^2 & \sum_{N(x)}w_i(x)^2(x_i-x) \cr 
       \sum_{N(x)}w_i(x)^2(x_i-x) & \sum_{N(x)}w_i(x)^2(x_i-x)^2 \cr}. $$ 
Hence its unconditional mean and covariance matrix are respectively 
$(a_0,b_0)$ and $\sigma^2\{S(x)^{-1}T(x)S(x)^{-1}+W_{0,x}^{-1}\}$. 
Consequently, 
$$P_0(x)\quad{\rm is\ unbiased\ for\ }\quad
        \sigma^2\{S(x)^{-1}T(x)+W_{0,x}^{-1}S(x)\}. \eqno(3.4)$$
Notice that $T(x)=S(x)$ when the uniform kernel is used in (1.3).
Note also that the natural trace statistic $d(x)'S(x)d(x)$
has expected value $2+{\rm Tr}\{W_{0,x}^{-1}S(x)\}$, 
if such uniform weights are used.  
These facts can be utilised in various ways to obtain empirical Bayes
estimates of parameters present in the $W_{0,x}^{-1}$ matrices,
as commented on further below. 

The arguments that led to (2.5) in the running one-parameter case
cannot be immediately generalised to the present running two-parameter case. 
The combined likelihood for the $k$ groups of local data becomes 
$$\prod_{x\in I}|\rho(x)|^{1/2}\,\sigma^{-\sum_{x\in I}s_0(x)}\,
        \exp\Bigl[-\half{1\over \sigma^2}\Bigl\{\sum_{x\in I}Q_0(x)
        +\sum_{x\in I}d(x)'S(x)\rho(x)d(x)\Bigr\}\Bigr], \eqno(3.5)$$
in terms of $\rho(x)=\{W_{0,x}+S(x)\}^{-1}W_{0,x}$. 
Taking partial derivatives to find the maximum here one ends up with 
$\tilda\rho(x)^{-1}=(1/\tilda\sigma^2)d(x)d(x)'S(x)$,
giving a local estimate of $\rho(x)^{-1}=I+W_{0,x}^{-1}S(x)$
in good agreement with (3.4). 
The difficulty is that the estimator is a rank 1 matrix, 
and there is in general no unique maximand $\tilda\rho(x)$.
Estimating $\sigma$ is less difficult. 
By writing $\rho(x)=r_xB_x$ for a $B_x$ matrix with determinant 1 
one can maximise over $\sigma$ and the $k$ values of $r_x$, to find 
$$\tilda\sigma^2=\sum_{x\in I}Q_0(x)
        \Big/\sum_{x\in I}\{s_0(x)-2\}. \eqno(3.6)$$
It is also easy to find Bayes estimates of $\sigma$ under 
a Gamma prior for $1/\sigma^2$. 
As in Section 2 a nice feature is that the $\sigma$ estimate 
is independent of start curve and of the specific forms used for $\rho(x)$. 

As similarly discussed in Section 2.2 satisfactory solutions 
are obtained by smoothing 
$$\rho(x)^{-1}=I+W_{0,x}^{-1}S(x) \quad {\rm against} \quad 
        (1/\tilda\sigma^2)d(x)d(x)'S(x) \eqno(3.7)$$
in suitable ways. For example, simple averaging suggests  
using the inverse of $(1/\tilda\sigma^2)\bar P_0$ as start curve weight, 
where $\bar P_0=k^{-1}\sum_{x\in I}d(x)d(x)'S(x)$, that is, 
$$\mtrix{\hatt m(x) \cr \hatt b(x)}
=\tilda\sigma^2\bar P_0^{-1}\mtrix{m_0(x) \cr m_0'(x)}
  +(I-\tilda\sigma^2\bar P_0^{-1})
        \mtrix{\tilda m(x) \cr \tilda b(x)}. \eqno(3.8)$$
The weight matrices are truncated to give weights in $[0,1]$ 
for both level and slope, if necessary. 
This is the natural extension of method (2.7).  

Other suitable solutions emerge by modelling $W_{0,x}$ parametrically
and then either use regression based on (3.6) 
or maximising (3.5) under this constraint. 
There does not appear to be a canonically unique way of doing this,
and several options may be considered.
An example is given in Section 4 where 
prior considerations suggest using $W_{0,x}=w_0A_x$ 
in terms of a single parameter times a known matrix function. 
A solution can be given much as in (2.8). 
Another example could be to model level and slope as 
locally independent and with $W_{0,x}={\rm diag}\{w_ar_a(x),w_br_b(x)\}$,
in terms of known functions $r_a(x)$ and $r_b(x)$ 
decided on by prior considerations. 
Again the combined likelihood (3.5) can be maximised 
with respect to $w_a$ and $w_b$, or a suitable regression analysis 
can produce estimates, based on 
$$d(x)'S(x)d(x)/\tilda\sigma^2
\simeq 2+w_a^{-1}s_0(x)/r_a(x)+w_b^{-1}s_2(x)/r_b(x). $$        
Presumably the exact specification of these parameters 
is of secondary importance compared to the specification of 
the start function $m_0$ or its prior distribution.

\subsection
{\csc 3.4. Accounting for uncertainty in the start curve.}
This quite crucial ingredient can be discussed 
very much as in Section 2.3. Again a feasible solution is to
take a parametric $m_0(x,\xi)$ as starting point, 
place a prior $\pi_0(\xi)$ on these background parameters, 
compute the exact or approximate posterior $\pi_0(\xi\midd{\rm all\ data})$,
leading in the end to 
$$\eqalign{
\mtrix{\hatt m(x) \cr \hatt b(x)}
=&\int\{\hatt W_{0,x}(\xi)+S(x)\}^{-1}
\Bigl\{\hatt W_{0,x}(\xi)\mtrix{m_0(x,\xi) \cr m_0'(x,\xi)} \cr
&\qquad\qquad 
+S(x)\mtrix{\tilda m(x) \cr \tilda b(x)}\Bigr\}
        \pi_0(\xi\midd{\rm all\ data})\,\d \xi, \cr} \eqno(3.9)$$
for example through simulation. 
The final Bayes estimator is $\hatt m(x)$. 

\bigskip
{\bf 4. A particular construction.} 
This section is meant to illustrate the general scheme of Section 3, 
while tending more carefully to some of the technical details. 
Suppose the start curve is simply a linear 
$m_0(x,\xi)=\xi_1+\xi_2(x-\bar x)$, 
where $\bar x$ is the average of the $x_i$s. 
The linear structure will be utilised to generate 
`the 100 likely start curves' that are  
needed to find the final Bayes estimate, 
as well as to suggest a structure for and then estimates 
of prior precision parameters. 
A generalisation is noted in Section 4.3,
while an attempt at making an automatic empirical Bayesian 
curve estimator is discussed in Section 4.4. 

\subsection
{\csc 4.1. Obtaining prior parameters.}
Imagine that the Bayesian's prior opinion about $\xi_1$ and $\xi_2$ 
is based on a previous data set with characteristics similar to the
present one. He would then have uncorrelated estimates 
$\xi_1^*$ and $\xi_2^*$ with variances respectively 
$n_0^{-1}\sigma_0^2$ and $n_0^{-1}\sigma_0^2/(v^*)^2$, say, 
in terms of prior sample size $n_0$ and prior sample 
$x$-variance $(v^*)^2$. 
This statistician would use prior variance 
$w_0^{-1}\{1+(x-\bar x)^2/v^2\}$ for $a=\xi_1+\xi_2(x-\bar x)$, 
prior variance $w_0^{-1}/v^2$ for $b=\xi_2$, with 
prior covariance $w_0^{-1}(x-\bar x)/v^2$ between the two,
in terms of the constant $w_0^{-1}=\sigma_0^2/n_0$
and the sample variance $v^2=n^{-1}\sumin(x_i-\bar x)^2$ for $x_i$s. 
This suggests using prior covariance matrix of the form 
$\sigma^2W_{0,x}^{-1}$ for the local $(a,b)$, where  
$$\eqalign{
W_{0,x}^{-1}&=w_0^{-1}\mtrix{1+(x-\bar x)^2/v^2 &(x-\bar x)/v^2 \cr
        (x-\bar x)/v^2 &1/v^2 \cr}, \cr
{\rm or}\quad W_{0,x}&=w_0\mtrix{1 & -(x-\bar x) \cr 
        -(x-\bar x) & v^2+(x-\bar x)^2}. \cr} \eqno(4.1)$$ 
This is accordingly a situation where prior knowledge has led
to a specific parametric form of the prior precision matrices,
say $W_{0,x}=w_0A_x$ or $W_{0,x}^{-1}=w_0^{-1}A_x^{-1}$ 
with known matrix function $A_x$. To estimate the $w_0$ quantity, 
one option is to maximise the combined marginal likelihood for 
the local data sets, that is, maximise 
$$\sum_{x\in I}\log{|w_0A_x|\over |w_0A_x+S(x)|}
-{1\over \tilda\sigma^2}\sum_{x\in I}
d(x,\xi_1,\xi_2)'\{w_0^{-1}A_x^{-1}+S(x)^{-1}\}^{-1}d(x,\xi_1,\xi_2) $$
with respect to $w_0$, which is an easy numerical task. Here 
$$d(x,\xi_1,\xi_2)=\mtrix{\tilda m(x)-\xi_1-\xi_2(x-\bar x) \cr 
        \tilda b(x)-\xi_2}, $$
with $\tilda m(x)$ and $\tilda b(x)$ as in the LL method (1.4). 
Alternatively one may use the structure 
$$d(x,\xi_1,\xi_2)'S(x)d(x,\xi_1,\xi_2)/\tilda\sigma^2
        =2+w_0^{-1}{\rm Tr}\{A_x^{-1}S(x)\}+{\rm error\ term}, $$
for several chosen values of $x$, 
assuming uniform weights in (1.3),   
to estimate $w_0^{-1}$ by a suitable regression analysis.
This also invites a graphical check on the 
`prior knowledge assumptions' that led to the $w_0A_x$ structure. 
Either way a Bayes-empirical-Bayes estimator 
has been constructed, of the form 
$$\eqalign{
\mtrix{\hatt m(x,\xi_1,\xi_2) \cr \hatt b(x,\xi_1,\xi_2)}
&=\{\hatt w_0(\xi_1,\xi_2)A_x+S(x)\}^{-1} \cr
&\times
\Bigl\{\hatt w_0(\xi_1,\xi_2)A_x\mtrix{\xi_1+\xi_2(x-\bar x) \cr \xi_2}
        +S(x)\mtrix{\tilda m(x) \cr \tilda b(x)}\Bigr\}, \cr}\eqno(4.2)$$
for each possible start curve $\xi_1+\xi_2(x-\bar x)$. 

\subsection
{\csc 4.2. Posterior distribution for start curve.} 
Next turn to the approximate distribution of start curves
given all data. The maximum likelihood estimators are 
$\tilda\xi_1=\bar y$ and $\tilda\xi_2=n^{-1}\sumin(x_i-x)y_i/v^2$.
The (2.10) matrix becomes 
$$\tilda V=\mtrix{
n^{-1}\sumin(y_i-\tilda y_i)^2, 
        &n^{-1}\sumin(y_i-\tilda y_i)^2(x_i-\bar x)/v^2 \cr
n^{-1}\sumin(y_i-\tilda y_i)^2(x_i-\bar x)/v^2, 
        &n^{-1}\sumin(y_i-\tilda y_i)^2(x_i-\bar x)^2/v^4 \cr}, $$
where $\tilda y_i$ is the fitted value $\tilda\xi_1+\tilda\xi_2(x_i-\bar x)$.
Only under the additional assumption that the underlying $m(\cdot)$ 
curve is actually linear is $\tilda V$ close to the familiar
$\tilda\sigma^2\,{\rm diag}\{1,1/v^2\}$.  
The general result explained in Section 2.3 implies that 
$(\xi_1,\xi_2)$ given the data information is approximately 
a binormal, centred at $(\tilda\xi_1,\tilda\xi_2)$ and with
covariance matrix $\tilda V/n$. 

The final estimator is reached as follows.
Simulate say 100 values of $(\xi_1,\xi_2)$ from the posterior
distribution just described. We may think of this as a way
of generating 100 likely start curves. 
For each such curve, compute the (4.2) estimate. 
In the end compute the average of the 100 curves 
$\hatt m(x,\xi_1,\xi_2)$. 

\def\zdot{z^*}
\subsection
{\csc 4.3. General linearly structured start curve.}
Assume that the start curve is parametrised as 
$m_0(x,\xi)=\xi_1+\xi_2g_2(x)+\cdots+\xi_pg_p(x)=\xi'z(x)$,
where $z(x)$ is the $p$-vector $(1,g_1(x),\ldots,g_p(x))'$. 
Its derivative is $m_0'(x,\xi)=\xi'\zdot(x)=\sum_{j=1}^p\xi_jg_j'(x)$.
Let $\tilda K=\{n^{-1}\sumin z(x_i)z(x_i)'\}^{-1}$.  
The same reasoning as above indicates that a reasonable 
structure for the covariance matrix of $(a,b)$ at $x$ is 
$$W_{0,x}^{-1}=w_0^{-1}\mtrix{z(x)'\tilda K z(x) &z(x)'\tilda K\zdot(x) \cr
z(x)'\tilda K\zdot(x) &\zdot(x)'\tilda K\zdot(x)}. $$
The scheme is otherwise quite similar to that above;
decide on a strategy to determine the $w_0$ parameter,
say $\hatt w_0(\xi)$, making it possible to arrive at 
a $\hatt m(x,\xi)$ as in (4.2) with a single algorithm, 
for each given $m_0(x,\xi)$. 
Then put up the $\tilda V/n$ matrix as per (2.10),
and finally compute the average of 100 curves $\hatt m(x,\xi)$.  

\subsection
{\csc 4.4. Completely automatic empirical Bayesian regression.} 
There are clearly many possible schemes to follow. 
This is inherent in the Bayesian perspective; 
each new application is different from previous ones and 
serious considerations are required 
to elicit the particular prior process, or at least its form. 
It is nevertheless useful to work out one or more methods 
that are completely automatic, in the sense of depending only
on the data and not on specification of subjective parameters, 
at the price of being pragmatically empirical Bayesian 
rather than strictly Bayesian. 
Indeed the two standard estimators (1.2) and (1.4) 
can both be given such interpretations, using flat priors 
for the local parameters. Less drastic simplifications 
emerge by letting the data lead to a suitable linear parametric form
$m_0(x,\xi)$, and then employ methods described in the preceding
subsections. 

One particular construction is as follows. 
Use a `delete-knot regression spline' method to approximate 
the curve with a function of the spline form 
$m_0(x,\xi)=\sum_{j=1}^p\xi_jg_j(x)$,
where $g_j(x)=\{(x-k_j)^+\}^3$, and the knots are 
placed at $k_1<\cdots<k_p$. The method described in 
Breiman and Peters (1992, Section 2.3) is automatic and 
succeeds in selecting a rather small number $p$ of such 
well-placed knots. Then go through the method of the previous subsection.  

\bigskip
{\bf 5. Other local regression schemes.}
This section briefly develops and discusses 
further specialisations of the general 
Bayesian local regression strategy. 
The apparatus is applied to other kinds of models in Section 7. 

\subsection
{\csc 5.1. Local polynomial Bayesian regression.}
It is clear that the methods and calculations of Section 3 
can be generalised to for example local quadratic or cubic regression,
without serious difficulties. 
The local quadratic model uses $m(t)=a+b(t-x)+c(t-x)^2$,
and the estimator is $\E\{a\midd{\rm local\ data}\}$.
Choosing the local polynomial order must be seen in 
connection with the choice of local data window width, see Section 8.3.  
Allowing three or four local parameters 
would typically require somewhat larger data windows than 
for the one- and two-parameter methods of Sections 2 and 3. 
The method of splines uses local cubic polynomials,
but the pieces are knotted together in a different way than with 
the present setup. 

\subsection
{\csc 5.2. Start curve function times correction.}
Suppose $m_0(t)$ is some initial description,
perhaps containing `global' parameters, and model
the true curve locally as $m(t)=m_0(t)a$ for $t\in N(x)$,
with a local correction factor $a$ not far from 1. 
If the local correction factor $a$ is modelled as 
$\normal\{1,\sigma^2/w_0\}$, then the Bayes solution becomes 
$$\hatt m(x)=m_0(x){w_0+u_0(x)\tilda a(x)\over w_0+u_0(x)}, $$
where $u_0(x)=\sum_{N(x)}w_i(x)m_0(x_i)^2$ and 
$\tilda a(x)=\sum_{N(x)}w_i(x)m_0(x_i)y_i/u_0(x)$. 
It is also of interest to note that if weights 
$\bar K(h^{-1}(x_i-x))m(x)^2/m(x_i)^2$ are used instead of $w_i(x)$,
that is, coming from the somewhat altered kernel 
$\bar K(z)m(x)^2/m(x+hz)^2$, then the method is a relative of 
one recently developed in Hjort and Glad (1994). 

An empirical Bayes scheme must be devised to estimate $w_0$,
for example utilising $P_0(x)=u_0(x)\{\tilda a(x)-1\}^2$ at various $x$s. 
Further variants emerge when the local model is 
$m(t)=m_0(t)\{a+b(t-x)\}$ for $t$ near $x$.

\subsection
{\csc 5.3. Other base densities in the local likelihood.} 
Our kernel smoothed local likelihood has been of the form 
$\prod_{N(x)}[\sigma^{-1}g(\sigma^{-1}\{y_i-a-b(x_i-x)\})]^{w_i(x)}$,
with $g$ equal to the standard normal. 
Other densities $g$ can be used as well, without seriously 
disturbing the general method, apart from more bothersome
numerical computations. An adaptive version of the method 
would be to stick in an estimate of $g$ based on 
residuals $\{y_i-\tilda m(x_i)\}/\tilda\sigma$. 
Analysis in Section 8.4 suggests 
that the choice of $g$ is not crucial to the final results, 
so we might as well stick to the convenient normal.  

\subsection
{\csc 5.4. Additional context in the prior.}  
Some of the development in Section 2.2 used the 
assumption that the local constant levels at the $k$
midpoint positions $x_{0,1}<\cdots<x_{0,k}$, 
say $a_1,\ldots,a_k$, were taken independent in their joint prior distribution.
One may also introduce an element of smoothness or context in the
prior, by modelling these as positively correlated, say 
${\bf a}\sim\normal_k\{{\bf a}_0,\sigma^2T_0^{-1}\}$. 
In such a case, 
$$\mtrix{\hatt m(x_{0,1}) \cr \vdots \cr \hatt m(x_{0,k})}
=(T_0+D)^{-1}T_0\mtrix{m_0(x_{0,1}) \cr \vdots \cr m_0(x_{0,k})}
+(T_0+D)^{-1}\mtrix{s_0(x_{0,1})\tilda m(x_{0,1}) \cr \vdots \cr 
        s_0(x_{0,k})\tilda m(x_{0,k})}, $$
where $D$ is the diagonal matrix with elements 
$(s_0(x_{0,1}),\ldots,s_0(x_{0,k}))$
(assuming, for simplicity, that uniform weights are used in (1.3)). 
The arguments that led to (2.5) and some of its later relatives
used a diagonal $T_0$ matrix. In the present setup 
the estimators again shrink the NW estimator towards the start curve
values $m_0(\cdot)$, but in addition neighbouring estimates 
are pushed towards each other, with a strength determined 
from the correlation structure in the prior. 
This is indeed sensible, and would not be very different 
from what happens for the method of splines,
which can be formulated in terms of a Gau\ss ian prior process 
for $m(\cdot)$ with positive correlations.  
We still prefer the simpler independence framework, however.
Context and smoothness is accounted for in any case 
through the use of a smooth start curve $m_0(\cdot)$ 
and direct signals from the data themselves. 
A correlation model is somewhat bothersome, 
not so much regarding the Bayes estimates above,
but by requiring an explicitly modelled correlation structure 
and a more complicated empirical Bayes scheme to estimate its parameters. 

\def\bfx{{\bf x}}
\def\bfb{{\bf b}}
\def\bft{{\bf t}}
\bigskip
{\bf 6. Several covariates.}
Nonparametric regression is difficult in higher dimensions.
It is easy to formally generalise many of the one-dimensional methods
to $d$ covariates, 
including the NW method of (1.2) and the LL method of (1.4), 
but the curse of dimensionality leaves 
most neighbourhoods too empty to give good precision,
and the convergence rate becomes increasingly unfavourable 
when $d$ grows. Successful methods, if say $d\ge 3$,
are typically those that look for lower-dimensional structure
in suitable ways. 
Friedman and Stuetzle (1981) develop a projection pursuit method. 
Cleveland and Devlin (1988) and Cleveland, Grosse and Shyu (1991) 
discuss multidimensional versions of the popular 
local regression method `lowess'. 
Hjort and Glad (1994) propose and analyse an estimator that 
corrects nonparametrically on an initial parametric descriptor. 
Hastie and Tibshirani (1990) discuss 
models and methods that approximate the real regression
surface with one with a simpler additive structure, 
and also (in their Ch.~6) give a Bayesian discussion of one version.
Here the apparatus of previous sections 
is extended to the $d$-variate case. 
Frank and Friedman (1993) study partial least squares methods 
and also point out Bayesian connections. 
The Bayesian methods may turn out to be particularly 
fruitful in the multivariate case in view of 
the difficulties that standard methods have there. 

The extension is quite straightforward in that the
necessary linear algebra is very similar to that developed in Section 3,
at least as concerns the structure of the Bayes estimator 
and supplementing empirical Bayes methods to obtain 
prior precision parameters. 
The model is that $y_i=m(\bfx_i)+\eps_i$ 
for a smooth surface $m(\cdot)$ 
in terms of say $\bfx_i=(x_{i,1},\ldots,x_{i,d})'$ for individual $i$, 
and i.i.d.~error terms $\eps_i$ with mean zero and variance 1.
The local model is 
$$m(t_1,\ldots,t_d)=a+b_1(t_1-x_1)+\cdots+b_d(t_d-x_d)
        =a+\bfb'(\bft-\bfx) 
        \quad {\rm for\ }\bft\in N(\bfx), $$ 
a suitably defined neighbourhood around $\bfx$.
Write 
$$Q(\bfx,a,\bfb)=\sum_{N(\bfx)}\{y_i-a-\bfb'(\bfx_i-\bfx)\}^2w_i(\bfx)
=Q_0(\bfx)+\mtrix{a-\tilda a \cr \bfb-\tilda\bfb}'S(\bfx) 
\mtrix{a-\tilda a \cr \bfb-\tilda\bfb}, $$
where $\tilda a,\tilda b_1,\ldots,\tilda b_d$ minimise $Q(x,a,\bfb)$,
and where 
$$S(\bfx)=\mtrix{\sum_{N(\bfx)}w_i(\bfx) 
        &\sum_{N(\bfx)}w_i(\bfx)(\bfx_i-\bfx)' \cr
\sum_{N(\bfx)}w_i(\bfx)(\bfx_i-\bfx)
        &\sum_{N(\bfx)}w_i(\bfx)(\bfx_i-\bfx)(\bfx_i-\bfx)' \cr}. $$
Let $(a,\bfb)$ be given a multinormal prior 
centred at $(a_0,\bfb_0)$, where $a_0=m_0(\bfx)$ 
and $\bfb_0$ has components $b_{0,j}=\dell m_0(\bfx)/\dell x_j$, 
determined from a suitable start surface $m_0(\bfx)$, 
and with a covariance matrix $\sigma^2W_{0,\bfx}^{-1}$.
Then results (3.1) and (3.2) hold
with only notational differences, and in particular this
defines the Bayes solution $\hatt m(\bfx)=\hatt a$. 

A successful version of this scheme,
especially with a respectable $d$, would typically require 
a good parametric modelling of the prior precision matrix $W_{0,\bfx}$
and then a suitable empirical Bayes method to infer
its parameters. To this end note that equation (3.4),
minorly modified, is valid. This also means that 
the appropriate analogue of (3.8) should be a good estimator 
in the present $d$-dimensional case. 
And as in previous cases in this paper this $m_0$-dependent estimator 
should be averaged with respect to the posterior distribution 
of the start surface.  

\bigskip
{\bf 7. Local linear Bayes methods for Poisson regression
and other regression models.} 
To illustrate that the general empirical-hierarchical-Bayesian
programme can be used in many other regression type models
we first consider the Poisson regression case in some detail
and then give brief pointers to still other areas of application.

\subsection
{\csc 7.1. Local inference for Poisson regression.} 
Let $y_i\midd x_i$ be a Poisson variable with mean parameter $m(x_i)$.
The task is to estimate the mean function $m(x)$.
Consider the local level model where $m(t)=a$ 
for $a\in N(x)=[x-\half h,x+\half h]$. 
The local likelihood becomes 
$$\prod_{N(x)}\Bigl\{e^{-a}a^{y_i}/(y_i!)\Bigr\}^{w_i(x)} 
=a^{\sum_{N(x)}w_i(x)y_i}
  \exp\Bigl\{-a\sum_{N(x)}w_i(x)\Bigr\}\Big/\prod_{N(x)}(y_i!)^{w_i(x)}. $$
Suppose there is some prior estimate of the form 
$m_0(x)=m_0(x,\xi)$, in terms of suitable `global' parameters $\xi$. 
If $a$ is given a Gamma prior with parameters $\{w_0m_0(x,\xi),w_0\}$,
that is, with mean value equal to $m_0(x,\xi)$  
and variance equal to $m_0(x,\xi)/w_0$, then 
$$a\midd{\rm local\ data},\xi\sim{\rm Gamma}
\Bigl\{w_0m_0(x,\xi)+\sum_{N(x)}w_i(x)y_i,\,
        w_0+\sum_{N(x)}w_i(x)\Bigr\}. $$
The Bayes estimator, conditional on the 
start curve $m_0(\cdot,\xi)$, becomes 
$$\hatt m(x,\xi)
=\E\{a\midd{\rm local\ data},\xi\}
={w_0\over w_0+s_0(x)}\,m_0(x,\xi)
        +{s_0(x)\over w_0+s_0(x)}\,\tilda m(x), \eqno(7.1)$$
where $s_0(x)=\sum_{N(x)}w_i(x)$ as before and 
$\tilda m(x)=\sum_{N(x)}w_i(x)y_i/s_0(x)$ 
is the natural frequentist estimate 
(maximum local likelihood estimate under the constant level model). 
This is also the Bayes solution under the natural 
noninformative prior (where $w_0$ tends to zero). 

As in Sections 2 and 3 empirical Bayes methods can be set up
to estimate the prior precision parameter $w_0=w_{0,x}$ at $x$. 
The marginal distribution of local data ${\cal D}(x)$,
still conditional on $w_{0,x}$ and the start curve, 
can be worked out to be 
$$\eqalign{
&{\Gamma(w_{0,x}m_0(x,\xi)+s_0(x)\tilda m(x))
   \over \Gamma(w_{0,x}m_0(x,\xi))
        s_0(x)^{s_0(x)\tilda m(x)}\prod_{N(x)}(y_i!)^{w_i(x)}} \cr
&\qquad\qquad\qquad\qquad
\times\Bigl\{{w_{0,x}\over w_{0,x}+s_0(x)}\Bigr\}^{w_{0,x}m_0(x,\xi)}
\Bigl\{{s_0(x)\over w_{0,x}+s_0(x)}\Bigr\}^{s_0(x)\tilda m(x)}. \cr}$$
The maximum likelihood estimator can be computed from this,
and is a suitable function of the sufficient statistic 
$\tilda m(x)$. It is simpler and perhaps equally reliable to 
utilise the easily obtainable facts that $\tilda m(x)$ 
has mean value $m_0(x,\xi)$ and variance 
$m_0(x,\xi)\{t_0(x)/s_0(x)^2+1/w_{0,x}\}$,
where $t_0(x)$ again is $\sum_{N(x)}w_i(x)^2$. 
This leads to forming 
$$P_0(x,\xi)={s_0(x)\over m_0(x,\xi)}\{\tilda m(x)-m_0(x,\xi)\}^2
\quad {\rm with\ mean\ value\ }
{t_0(x)\over s_0(x)}+{s_0(x)\over w_{0,x}}, \eqno(7.2)$$
giving a prior precision estimate $\hatt w_{0,x}=\hatt w_{0,x}(\xi)$. 
This leads to the Bayes-empirical-Bayes estimator 
$$\eqalign{m^*(x,\xi)
&={\hatt w_{0,x}(\xi)m_0(x,\xi)+s_0(x)\tilda m(x)
        \over \hatt w_{0,x}(\xi)+s_0(x)} \cr
&={1\over 1+P_0(x,\xi)-t_0(x)/s_0(x)}m_0(x,\xi)
        +{P_0(x,\xi)-t_0(x)/s_0(x)
        \over 1+P_0(x,\xi)-t_0(x)/s_0(x)}\tilda m(x). \cr} \eqno(7.3)$$
This is quite similar to the (2.6) estimator. 
The structure is simplest for the uniform kernel, 
where $t_0(x)/s_0(x)=1$,
and otherwise this ratio is close to $R_K/K(0)$,  
which for example is equal to 0.80 for the optimal 
Jepanetsjnikoff kernel (see Section 8.3). 
The reasoning that led to estimator (2.7) gives 
$$m^*(x,\xi)=\bar P_0(\xi)^{-1}m_0(x,\xi)
        +\{1-\bar P_0(\xi)^{-1}\}\tilda m(x), \eqno(7.4)$$
where $\bar P_0(\xi)=k^{-1}\sum_{x\in I}P_0(x,\xi)$. 
Again the weights are truncated to the unit interval. 
Similarly an analogue of estimator (2.8) can easily be constructed. 
The key step is to model the $w_{0,x}$ suitably as a function of $x$. 
Estimator (7.3) does not use any model at all for how
$w_{0,x}$ changes, and is quite nonparametric on this account.
Estimator (7.4) and suitable analogues of estimator (2.8)
use parametric models for $w_{0,x}$; 
see the discussion that led to (2.7) and (2.8). 

All this happened conditional on a start curve function $m_0(x,\xi)$.
As in previous sections a natural two-stage Bayesian way to 
cope with uncertainty in the specification of the start curve
is to place a prior on the parameters $\xi$, then compute 
the exact or approximate posterior distribution 
$\pi_0(\xi\midd{\rm all\ data})$. 
A natural scenario is a log-linear start curve model,
say $m_0(x,\xi)=\exp(\xi_1+\xi_2x)$ 
or the more general 
$\exp\{\xi'z(x)\}=\exp\{\xi_1+\xi_2g_2(x)+\cdots+\xi_pg_p(x)\}$.
The posterior distribution of $\xi$ is approximately 
a multinormal, centred at the maximum likelihood estimate $\tilda\xi$
and with a covariance matrix of form $\tilda V/n$, where in fact 
$$\tilda V=\Bigl(n^{-1}\sumin e^{\tilda\xi'z_i}z_iz_i'\Bigr)^{-1}
\Bigl\{n^{-1}\sumin(y_i-e^{\tilda\xi'z_i})^2z_iz_i'\Bigr\}
\Bigl(n^{-1}\sumin e^{\tilda\xi'z_i}z_iz_i'\Bigr)^{-1}. $$
This can be shown using methods of Hjort and Pollard 
(1994, Sections 3 and 4). The form for $\tilda V$ given here 
assumes that $y_i$ given $x_i$ is really a Poisson,
but does not assume that the mean function is of any 
particular form. The final estimator for $m(x)$ is of the form 
$$\eqalign{\hatt m(x)
&=\E\big[\E\{a\midd{\rm local\ data},\xi\}\midd{\rm all\ data}\bigr] \cr
&=\int {\hatt w_{0,x}(\xi)m_0(x,\xi)+s_0(x)\tilda m(x)
        \over \hatt w_{0,x}(\xi)+s_0(x)}
        \pi_0(\xi\midd{\rm all\ data})\,\d \xi. \cr}\eqno(7.5)$$

\subsection
{\csc 7.2. Local log-linear analysis.}
This time take $m(t)=a\exp\{b(t-x)\}$ as the local model around $x$.
Let $a$ be a Gamma with parameters $\{w_0m_0(x,\xi),w_0\}$ as above,
and let $b$ have some prior $\pi(b)$.  
Analysis like in the previous subsection shows that 
$a$ given local data and $b$ is an updated Gamma, with 
$$\E\{a\midd{\rm local\ data},b\}
={w_0m_0(x,\xi)+\sum_{N(x)}w_i(x)y_i
        \over w_0+\sum_{N(x)}w_i(x)\exp\{b(x_i-x)\}}, $$
and the Bayes estimator is the average of this over the
distribution of $b$ given local data,
$$\eqalign{\hatt m(x,\xi)
&=\E\{m(x)\midd{\rm local\ data}\} \cr
&=\int {w_0m_0(x,\xi)+\sum_{N(x)}w_i(x)y_i
        \over w_0+\sum_{N(x)}w_i(x)\exp\{b(x_i-x)\}}
        \pi(b\midd{\rm local\ data})\,\d b. \cr}$$
The posterior density for $b$ can be shown to be of the form 
$${\rm const.}\,\pi(b){\exp\{bnh^3u_n(x)/k_0\}
\over \{w_0+nhf_n(x)/k_0+nh^3v_n(x,b)/k_0\}
        ^{w_0m_0(x,\xi)+nhf_n(x)\tilda m(x)/k_0}}, $$
where $u_n(x)=n^{-1}h^{-3}\sum_{N(x)}K(h^{-1}(x_i-x))(x_i-x)y_i$ 
essentially estimates $(mf)'$, and where 
$v_n(x,b)=n^{-1}h^{-3}\sum_{N(x)}K(h^{-1}(x_i-x))[\exp\{b(x_i-x)\}-1]$ 
essentially estimates $f''+2bf'+b^2f$. 
Normal approximations of interest can be worked out based on this,
but will not be pursued here. As in the previous subsection 
one must next supply estimates of the $w_0=w_{0,x}$ values 
and in the end average the Bayes-empirical-Bayes estimate $m^*(x,\xi)$ 
over the posterior distribution of $\xi$ given all data.  
 
\subsection
{\csc 7.3. Start estimate times local correction.}
This time take $m(t)=m_0(t,\xi)a$ for $a\in N(x)=[x-\half h,x+\half h]$ 
as the local model, where $a=a_x$ is thought of 
as the local multiplicative correction factor
to the start curve $m_0(x,\xi)$.  
The local likelihood becomes proportional to 
$$\eqalign{
&\prod_{N(x)}\Bigl[\exp\{-m_0(x_i,\xi)a\}
        \{m_0(x_i,\xi)a\}^{y_i}\Bigr]^{w_i(x)} \cr
&=\prod_{N(x)}m_0(x_i,\xi)^{y_iw_i(x)}\,
a^{\sum_{N(x)}w_i(x)y_i}
        \exp\Bigl\{-a\sum_{N(x)}w_i(x)m_0(x_i,\xi)\Bigr\}. \cr}$$
Under present circumstances it is appropriate to give 
$a$ a Gamma prior centred around 1, say with parameters $(w_0,w_0)$.
Then $a$ given local data and the background $\xi$ is seen to
be an updated Gamma, 
and the Bayes estimator is 
$$\hatt m(x,\xi)=\E\{m_0(x,\xi)a\midd{\rm local\ data},\xi\}
        =m_0(x,\xi){w_0+\sum_{N(x)}w_i(x)y_i
        \over w_0+\sum_{N(x)}w_i(x)m_0(x_i,\xi)}. $$
Note that the noninformative prior version of this gives the 
interesting estimator 
$$\bar m(x,\xi)=\sum_{N(x)}w_i(x)m_0(x,\xi)y_i\Big/
        \sum_{N(x)}w_i(x)m_0(x_i,\xi), $$
which is equal to the ordinary NW-type one only if the start curve
function is constant in the neighbourhood. 
It has exactly the same `start estimator times correction' structure 
as that of estimators worked with in Hjort and Glad (1994) 
provided the kernel $\bar K(z)m_0(x,\xi)/m_0(x,\xi+hz)$ is used.  

The programme is once more to estimate $w_0=w_{0,x}$ in 
a suitable fashion, for given $\xi$, ending up with 
a Bayes-empirical-estimator $m^*(x,\xi)$, and then 
averaging this over say 100 likely prior curves drawn with respect to
the posterior density of $\xi$ given all data. 
Helpful for the first step here is to use 
$$P_0(x,\xi)=\sum_{N(x)}m_0(x,\xi)\Bigl\{{\bar m(x,\xi)\over m_0(x,\xi)}
        -1\Bigr\}^2\simeq 1+w_{0,x}^{-1}\sum_{N(x)}m_0(x_i,\xi), $$
with uniform kernel weighting in (1.3). 

\subsection
{\csc 7.4. Local linear Bayes analysis of other regression models.}
The ideas and methods of this paper can be applied in many other 
regression situations, such as logistic regression 
and Cox regression in survival analysis. 
Yet another situation where similar methods can be put forward
is that of spatial interpolation of random fields;
the result would be a local version of the Bayesian Kriging 
method treated in Hjort and Omre (1994, Section 3). 

\bigskip
{\bf 8. Supplementing results and remarks.}

\subsection
{\csc 8.1. Stein-type estimation and risk function comparison.}
The following discussion is pertinent also for methods 
developed in the later sections, but to keep matters simple 
we consider the situation in Section 2, 
where the local levels $m(x)$ at $k$ positions were to be estimated. 
Let us also for simplicity 
employ the uniform kernel in (1.3) so that $t_0(x)$ and $s_0(x)$ 
are both equal to the number of data points falling inside $x\pm\half h$. 
Suppose the loss function involved is 
$$L(m,m^*)=\sum_{x\in I}\{m^*(x)-m(x)\}^2s_0(x). $$
The standard estimator $\tilda m(x)$ has risk function 
equal to the constant $k\sigma^2$. 
In view of estimators (2.7) and (2.8), let us try out 
$$m^*(x)=\tilda m(x)-c\{\tilda m(x)-m_0(x)\}/z,
\quad {\rm where\ }z=\sum_{x\in I}\{\tilda m(x)-m_0(x)\}^2s_0(x). $$
Its loss can be written 
$$L(m,m^*)=L(m,\tilda m)+c^2/z
-2c\sum_{x\in I}s_0(x)\{\tilda m(x)-m(x)\}\{\tilda m(x)-m_0(x)\}/z. $$
Using partial integration and properties of the normal distribution 
one sees that  
$$\eqalign{
&\E_m\{\tilda m(x)-m(x)\}q(\tilda m(x_{0,1}),\ldots,\tilda m(x_{0,k})) \cr
&\qquad\qquad
=\{\sigma^2/s_0(x)\}\E_m\{\dell/\dell\tilda m(x)\}
        \,q(\tilda m(x_{0,1}),\ldots,\tilda m(x_{0,k})). \cr}$$
This implies, with some calculations, that the risk of $m^*$ 
is equal to the risk of $\tilda m$ plus $\E_m\Delta$, 
where $\Delta=z^{-1}\{c^2-2c(k-2)\sigma^2\}$. The best value for $c$
is $c_0=(k-2)\sigma^2$, making the $\Delta$ function 
negative for all values. 
This shows that $\tilda m$ can be improved upon not only in a region
around a given prior estimate $m_0$, but uniformly, if only $k\ge3$. 
This is the Stein phenomenon, see for example Lehmann (1983, Ch.~4)
for discussion of this in a somewhat simpler framework. 
The development here suggests the estimator 
$$m^*(x)=\tilda m(x)-{(k-2)\tilda\sigma^2
        \over \sum_{x\in I}\{\tilda m(x)-m_0(x)\}^2s_0(x)}
        \{\tilda m(x)-m_0(x)\}, \eqno(8.1)$$
which is quite similar to (2.7). 

Further studies are needed in order to single out practical
versions of our schemes with good performance against 
traditional competitors. A simulation study along the lines
of Breiman and Peters (1992) could be carried out, 
using proposals as outlined in Section 4.4 above, for example. 

\subsection
{\csc 8.2. Alternative estimators for $\sigma$ and prior precision.}
Other estimators can also be developed for $\sigma$ than in (2.5). 
With a prior $\pi_0(\sigma)$ for the $\sigma$ parameter 
the simultaneous distribution of $\sigma$ and the combined 
neighbourhood data sets ${\cal D}(x)$ can be written down, 
following the expression that led to (2.5). 
A convenient choice is the Gamma prior with parameters say 
$(\half\alpha,\half\beta)$ for $\lambda=1/\sigma^2$,
with prior guess $\sigma_0^2=\beta/\alpha$ for $\sigma^2$.  
Then $\lambda$, given the collection of all the local data sets, 
is still a Gamma with updated parameters 
$\half\alpha+\half\sum_{x\in I}s_0(x)$
and $\half\beta+\half\sum_{x\in I}\{Q_0(x)+\rho(x)P_0(x)\}$. 
This leads to the Bayes estimator 
$$\hatt\sigma^2=\{\alpha+\sum_{x\in I}s_0(x)\}^{-1}
\Bigl[\beta+\sum_{x\in I}Q_0(x)+\sum_{x\in I}
        {w_{0,x}s_0(x)\over w_{0,x}+s_0(x)}
        \{\tilda m(x)-m_0(x)\}^2\Bigr]. $$
It is also of interest to note that if the unconditional distribution
of the $k$ local data sets is deduced, from the above by
integrating out $\sigma$, then its maximisers are exactly as in (2.5). 
The noninformative prior version of this is the one where
$\alpha$ and $\beta$ tend to zero, corresponding in fact to having 
a uniform prior for $\log\sigma$. 
One may also let the $w_{0,x}$ parameters tend to zero,
corresponding to a uniform prior for each of the $k$ local levels $a(x)$.
This invites $\sum_{x\in I}Q_0(x)/\sum_{x\in I}s_0(x)$,
which is quite similar to the one in (2.5). 
Yet other estimators of $\sigma$ are discussed in 
Hastie and Tibshirani (1990, Section 3.4).

We saw in (2.3) that the $\sigma$-conditional posterior 
distribution of the local constant $a$ was a normal with
variance proportional to $\sigma^2$. 
The real local posterior distribution of $a$ emerges by
integrating this normal with respect to the posterior distribution 
of $\sigma$. With the Gamma prior $(\half\alpha,\half\beta)$ 
for $1/\sigma^2$ used above some calculations show that 
$$a=m(x)\midd{\rm local\ data}\sim \hatt m(x)
        +\{w_{0,x}+s_0(x)\}^{-1/2}\hatt\sigma\,t_\nu, $$
where $t_\nu$ is a $t$ distribution with degrees of freedom equal
to $\nu=\alpha+\sum_{x\in I}s_0(x)$. With the noninformative 
prior on $\sigma$ and uniform weights we get a $t$ 
with $n$ degrees of freedom. This leads to pointwise Bayesian 
credibility bands for the $m(x)$ curve. 

\subsection
{\csc 8.3. Choosing kernel, bandwidth and order.}
The local likelihood $L_n(x,\beta,\sigma)$ of (1.5) 
with a uniform kernel has discontinuities in $x$ 
when the endpoints of the $x\pm\half h$
interval hit data points. This drawback is inherited for 
both Bayes estimators and maximum local likelihood estimators.
Necessary for continuity of these is continuity of 
the kernel function $\bar K$ in the full $[-\half,\half]$ 
support interval, in particular $\bar K(\pm\half)=0$ is required.  
Let $K$ be a probability density kernel with 
standard deviation $\sigma_K$ and $R_K=\int K^2\,\d z$,
and let $\bar K$ be related via (1.3). 
The approximate or asymptotic mean squared error of 
the maximum local likelihood estimator, say the LL estimator 
(1.4), can be expressed as $\{\sigma_KR_K\}^{4/5}$ times 
a factor depending on the unknown $m(\cdot)$ and then 
divided by $n^{4/5}$. The best kernel in this sense is
the one supported on $[-\half,\half]$ and minimising $\sigma_KR_K$.
This is the Jepanetsj\-nikoff kernel, and in terms of $\bar K$ 
this is $\bar K(z)=1-(2z)^2$ on $[-\half,\half]$ and zero outside. 

Choosing bandwidth and order of the local model is 
a nontrivial problem, chiefly related to the variance--bias balancing act. 
Methods and insight reached in the non-Bayesian 
local regression context are mostly relevant also in the  
present Bayesian framework, and a pragmatic view would 
be to let the non-Bayesians decide on $h$ and the local
polynomial order, as best they can for a given problem,
and then use the Bayesian methods developed here with the
chosen $h$ and order.  
Cleveland and Devlin (1988) use certain $M$-plots that 
resemble Mallows' $C_p$-statistics. 
Tibshirani and Hastie (1987) use Akaike's information criterion
(with a uniform kernel, and with symmetric nearest neighbours
windows); the so-called Bayesian information criterion of 
Schwarz (1978) could as easily be used. Fan and Gijbels (1992)
consider versions of `plug-in' methods to decide on $h$. 

Various Bayesian methods can also be developed, 
including fanciful ones that for a given order start with a prior 
for the $h$ or the $h_x$ process. 
An easier method, which is Bayesian in the sense that 
it can incorporate prior knowledge, 
is to work with the widest practical model, 
say the third order local regression 
$$m(t)=a+b(t-x)+c(t-x)^2+d(t-x)^3
        \quad {\rm for\ }t\in[x-\half h,x+\half h], $$
and model the local covariance matrix $W_{0,x}^{-1}$ 
so as to suitably penalise third and second order presence. 
The result would also resemble a special case of 
general smoothing-between-models estimators that are discussed 
generally in Hjort (1994c). 
Rather than developing such ideas  
we are content here to describe a natural 
local goodness of fit method, which also might be useful 
for the non-Bayesian local regression methods. 
For the running local line case (local polynomial order 1), let 
$$Q_0(x)=\sum_{N(x)}\{y_i-\tilda a(x)-\tilda b(x)(x_i-x)\}^2, $$
using uniform weights for simplicity.  
Under the hypothesis that the regression really is linear
in the $N(x)=x\pm\half h$ interval, and with normal residuals, 
$Q_0(x)/\sigma^2\sim\chi^2_{s_0(x)-2}$,
where $s_0(x)$ is the number of data points falling in $N(x)$ window.
A rough strategy is therefore to expand the $x\pm\half h$ window,
from a suitable minimum length $h_0(x)$ onwards, 
as long as $Q_0(x)/\tilda\sigma^2\le q(0.80,s_0(x)-2)$,
the upper 20\% point (for example) of the $\chi^2$ 
with $s_0(x)-2$ degrees of freedom.
A kurtosis correction can readily be supplied if the residuals 
are non-normal, and the chosen values of $h=h_x$ should be 
post-smoothed somewhat to give a smooth curve. 
There are obvious modifications of this method for 
local polynomial orders say 2 and 3
(in particular, with appropriate versions of $\tilda\sigma^2$ 
of (3.6), subtracting respectively 3 and 4 in the denominator). 
In the end the resulting estimators, say of order 1, 2, and 3, 
can be scrutinised and one of them could be selected 
from a suitable overall criterion.  

\subsection
{\csc 8.4. What is the kernel smoothed local likelihood aiming at?} 
Consider the local kernel smoothed likelihood $L_n(x,a,b)$ 
that was the starting point of Section 3. 
Indirectly its use hinges on considering  
the normal $\{a+b(t-x),\sigma^2\}$ model, say $f(y\midd t,a,b,\sigma)$, 
to be a relevant approximation to the real $f(y\midd t)$ density,  
for $t$ in the vicinity of a given $x$. 
To quantify this, note that 
$n^{-1}$ times the log-local likelihood tends to
$$\int \bar K(h^{-1}(t-x))f(t)\,
\Bigl\{\int f(y\midd x)\log f(y\midd t,a,b,\sigma)\,\d y\Bigr\}\,\d t, $$
by the law of large numbers. This shows that estimation based on
$L_n(x,a,b,\sigma)$ aims at approximating the real $f(y\midd t)$ as
well as possible in the sense of minimising 
the locally weighted distance 
$$\int_{N(x)} K_h(t-x)f(t)\,
        \Delta[f(\cdot\midd t),f(\cdot\midd t,a,b,\sigma)]\,\d t $$
between true and parametrically modelled distributions,
where $K_h$ is the scaled version $K_h(z)=h^{-1}K(h^{-1}z)$ and 
where $\Delta[f,g]$ is the Kull\-back--Leibler distance 
$\int f\log(g/f)\,\d y$.
In the present case, this can be seen to be the same as minimising 
$$\int_{N(x)} K_h(t-x)f(t)\bigl\{\log\sigma
+\half[\sigma_{\rm tr}^2+\{m(t)-a-b(t-x)\}^2/\sigma^2]\bigr\}\,\d t, $$
where $\sigma_{\rm tr}$ is the underlying true standard deviation
for $y_i-m(x_i)$. This shows that local linear modelling aims to provide the
best approximation $a_x+b_x(t-x)$ to the true $m(t)$ around $x$
in the sense of minimising $\int K_h(t-x)f(t)\{m(t)-a_x-b_x(t-x)\}^2\,\d t$.
This readily implies $a_x=m(x)+O(h^2m''(x))$. 
Also, the least false $\sigma^2$ aimed at is 
$${\sum_{x\in I}\int_{N(x)}K_h(t-x)f(t)[\sigma_{\rm tr}^2
        +\{m(x)-a_x-b_x(t-x)\}^2]\,\d t
\over \sum_{x\in I}\int_{N(x)}K_h(t-x)f(t)\,\d t}, $$
which is slightly bigger than the true variance parameter.
The difference is $O(h^2)$ and usually small. 
These results suggest that quite sensible best 
local approximating models are aimed for, and, in particular,
that the indirectly utilised normality assumption cannot matter much.
See Section 5.3 for non-normal local likelihoods. 


\subsection
{\csc 8.5. Avoiding the binning of data.} 
Several of the methods arrived at here depend at the outset 
on the particular binning of data into cells, 
with $\cup_{x\in I}{\cal D}(x)$.
Approximations of the quantities involved can be developed
to replace sums over $x\in I$ with sums over all data points, 
thus avoiding dependence on any particular binning.

\subsection
{\csc 8.6. Nearness of the two basic estimators.}
The following calculations indicate that estimators stemming from
the local constant model (Section 2) often can be viewed as simpler 
approximations to those stemming from the local linear model
(Sections 3 and 4). Those using the local $a+b(t-x)$ model 
are essentially performing an $O(h^2)$ debiasing operation on 
those using the simpler $a$ model. 

With notation as in Section 3.1, start out noting that the sizes of 
$s_0$, $s_1$ and $s_2$ depend on the underlying density of the $x$s, 
say $f$, and its first and second derivatives. 
For the symmetric probability density kernel $K$ used 
in connection with the weights $w_i(x)$, see (1.3), 
let $k_j=\int z^jK(z)\,\d z$ for $j\ge1$ and let $k_0=K(0)$. 
Consider
$$\eqalign{
f_n(x)&=n^{-1}h^{-1}\sumin K(h^{-1}(x_i-x)), \cr
g_n(x)&=n^{-1}h^{-3}\sumin (x_i-x)K(h^{-1}(x_i-x)), \cr
q_n(x)&=n^{-1}h^{-2}\sumin \{h^{-2}(x_i-x)^2-k_2\}K(h^{-1}(x_i-x)). \cr}$$ 
These functions are essentially estimates of 
$f(x)$, $f'(x)$ and $f''(x)$; 
indeed $f_n$ has mean $f+O(h^2)$, 
$g_n$ has mean $k_2f'+O(h^2)$, and 
$q_n$ has mean $\half(k_4-k_2^2)f''+O(h^2)$. And 
$$\eqalign{
s_0(x)&=nhf_n(x)/k_0, \cr 
s_1(x)&=nh^3g_n(x)/k_0, \cr 
s_2(x)&=nh^5q_n(x)/k_0+k_2nh^3f_n(x)/k_0. \cr} $$
In the typical large-sample analysis the smoothing parameter 
$h$ has to tend slowly to zero in order to achieve
consistent estimation of $f$, $f'$ and $f''$, 
in fact $nh^5\arr\infty$ is required here. 
This shows that the size of $s_0(x)$ is typically bigger than 
those of $s_1(x)$ and $s_2(x)$. 
Some analysis also shows that $\tilda m_{\rm LL}(x)$ 
essentially performs an $O(h^2)$ debiasing type `correction' on 
$\tilda m_{\rm NW}(x)$. This also says that the NW estimator 
can be seen as a first order approximation to the LL estimator, 
when $h$ is small. Corresponding remarks are valid for the 
Bayes estimators. 

\subsection
{\csc 8.7. Bayes estimation of regression derivatives.}
Suppose the first derivative $m'(x)$ is to be estimated. 
A natural method is to use local models of the form 
$m(t)=a+b(t-x)+c(t-x)^2$ around each given $x$, 
then carry out Bayesian estimation of the local parameters 
by conditioning on local data, and in the end use 
the $\tilda b=\tilda b_x$ component. 

\bigskip
{\bf 9. Conclusions.} 
We have described a general Bayesian/empirical Bayesian 
local regression method, comprising up to five steps (a)--(e),
as detailed in Section 1.3. 
There is a bewildering plethora of possible implementations 
of the general idea, as witnessed in Sections 2--4.
It is worth stressing that the class of local linear regression methods,
which has enjoyed increased popularity recently, emerges as 
the special case of the Bayesian programme which corresponds to
flat priors on the local parameters. 
The breadth of the spectrum of solutions is a consequence of 
the Bayesian paradigm; each new application could have a different prior
on the start curve and a different structure of the 
prior covariance matrix $\sigma^2W_{0,x}^{-1}$ (see Section~3).
In addition there are several ways of carrying out the empirical
Bayes step, that of estimating parameters in the $W_{0,x}$ matrices
from data. The methods of Section~4 end up as quite definite
proposals, though, in situations where the start curve can 
be parametrised linearly in basis functions. 
We have also made the point that the Bayes solutions have the potential
of outperforming classical methods, not only in the vicinity of 
prior guesses, but uniformly. 
Crucial factors for successful applications might include 
a good parametric representation of the start curve $m_0(x,\xi)$
and that of a good representation for the prior precision matrices 
$W_{0,x}$. Further study is needed in order to single out 
the best practical ways of doing local Bayesian regression.  

\bigskip
\centerline{\bf References} 

\parindent0pt
\baselineskip11pt
\parskip3pt
\medskip 

\ref{%
Box, G.E.P.~and Tiao, G.C. (1973).
{\sl Bayesian Inference in Statistical Analysis.}
Addison-Wesley, Menlo Park.}

\ref{%
Breiman, L.~and Peters, S. (1992).
Comparing automatic smoothers (a public service enterprise).
{\sl International Statistical Review} {\bf 60}, 271--290.} 

\ref{%
Cleveland, W.S. (1979). 
Robust locally weighted regression and smoothing scatterplots.
{\sl Journal of the American Statistical Association} 
{\bf 74}, 829--836.}

\ref{%
Cleveland, W.S.~and Devlin, S.J. (1988).
Locally weighted regression: 
An approach to regression analysis by local fitting. 
{\sl Journal of the American Statistical Association} 
{\bf 83}, 596--610.}

\ref{%
Cleveland, W.S., Grosse, E.~and Shyu, W.M. (1991).
Local regression models. 
In {\sl Statistical Models in S} (J.~Chambers and T.~Hastie, eds.), 309--376. 
Wadsworth, Pacific Grove, California.}

\ref{%
Erkanli, A., M\"uller, P.~and West, M. (1992).
Bayesian curve fitting using multivariate normal mixtures.
{\sl Biometrika}, to appear.}

\ref{%
Fan, J.~and Gijbels, I. (1992).
Variable bandwidth and local linear regression smoothers.
{\sl Annals of Statistics} {\bf 20}, 2008--2036.}


\ref{%
Ferguson, T.S., Phadia, E.G., and Tiwari, R.C. (1992).
Bayesian nonparametric inference.
In {\sl Current Issues in Statistical Inference: Essays in Honor of 
D.~Basu} (M.~Ghosh and P.K.~Patnak, eds.), 127--150.
IMS Lecture Notes \& Monograph Series {\bf 17}.}

\ref{%
Frank, I.E.~and Friedman, J.H. (1993).
A statistical view of some chemometrics regression tools
(with discussion).
{\sl Technometrics} {\bf 35}, 109--148.}

\ref{%
Friedman, J.H.~and Stuetzle, W. (1981).
Projection pursuit regression.
{\sl Journal of the American Statistical Association} 
{\bf 76}, 817--823.}

\ref{%
Hastie, T.J.~and Tibshirani, R.J. (1990).
{\sl Generalized Additive Models.}
Chapman and Hall, London.}

\ref{%
Hastie, T.J.~and Loader, C.R. (1993).
Local regression: Automatic kernel carpentry (with discussion).
{\sl Statistical Science} {\bf 8}, 120--143.}



\ref{%
Hjort, N.L. (1994a).
Dynamic likelihood hazard rate estimation.
To appear, hopefully, in {\sl Biometrika}.}

\ref{%
Hjort, N.L. (1994b).
Bayesian approaches to non- and semiparametric density estimation.
Invited paper presented at the Fifth Valencia International Meeting 
on Bayesian Statistics, to be published, hopefully, 
in {\sl Bayesian Statistics 5}.}

\ref{%
Hjort, N.L. (1994c).
Estimation in moderately misspecified models.
To appear, hopefully, in 
{\sl Journal of the American Statistical Association}.}

\ref{%
Hjort, N.L.~and Glad, I.K. (1994).
Nonparametric density estimation with a parametric start.
To appear, hopefully, in {\sl Annals of Statistics}.}

\ref{%
Hjort, N.L.~and Jones, M.C. (1994).
Locally parametric nonparametric density estimation.
To appear, hopefully, in {\sl Annals of Statistics}.}

\ref{%
Hjort, N.L.~and Omre, H. (1994).
Topics in spatial statistics (with discussion). 
To appear, hopefully, in {\sl Scandinavian Journal of Statistics}.}

\ref{%
Hjort, N.L.~and Pollard, D.B. (1994).
Asymptotics for minimisers of convex processes.
To appear, hopefully, in {\sl Annals of Statistics}.}

\ref{%
Lehmann, E.L. (1983).
{\sl Theory of Point Estimation.}
Wiley, New York.} 

\ref{%
Ruppert, D.~and Wand, M.P. (1994).
Multivariate locally weighted least squares regression.
{\sl Annals of Statistics}, to appear.}

\ref{%
Schwarz, G. (1978).
Estimating the dimension of a model.
{\sl Annals of Statistics} {\bf 6}, 461--464.} 
 
\ref{%
Scott, D.W. (1992).
{\sl Multivariate Density Estimation:
Theory, Practice, and Visualization.}
Wiley, New York.}

\ref{%
Silverman, B.W. (1985). 
Some aspects of the splines smoothing approach to nonparametric 
regression curve fitting (with discussion).
{\sl Journal of the Royal Statistical Society B} 
{\bf 46}, 1--52.}


\ref{%
Stone, C.J. (1977).
Consistent nonparametric regression (with discussion).
{\sl Annals of Statistics} {\bf 5}, 595--645.} 


\ref{%
Tibshirani, R.J.~and Hastie, T.J. (1987).
Local likelihood estimation.
{\sl Journal of the American Statistical Association} {\bf 82}, 559--567.}

\ref{%
Wahba, G. (1990).
{\sl Spline Models for Observational Data.}
SIAM, Philadelphia.}

\ref{%
Wand, M.P.~and Jones, M.C. (1994).
{\sl Kernel Smoothing.}
Chapman and Hall, London. To exist.}  

\ref{%
West, M., M\"uller, P.~and Escobar, M.D. (1994).
Hierarchical priors and mixture models, 
with application in regression and density estimation.
In {\sl Aspects of Uncertainty: A Tribute to D.V.~Lindley} 
(A.F.M.~Smith and P.R.~Freeman, eds.). Wiley, London.}

\bye